\documentclass[%
 reprint,
 amsmath,amssymb,
 aps,nofootinbib,
]{revtex4-1}
\usepackage{bm}% bold math
\PassOptionsToPackage{linktocpage}{hyperref}
\usepackage[hyperindex,breaklinks]{hyperref}
\usepackage{enumitem}
\usepackage{slashed}

\renewcommand{\theta}{\vartheta}

\usepackage{array}
\usepackage{mathtools}

\usepackage{etoolbox}

\begin{document} 

\title{Topological Origin of Chiral Symmetry Breaking in QCD and in Gravity}

\author{Gia Dvali} 
%\email{Georgi.Dvali@physik.uni-muenchen.de}
\affiliation{%
Arnold Sommerfeld Center, Ludwig-Maximilians-Universit\"at, Theresienstra{\ss}e 37, 80333 M\"unchen, Germany, 
}%
 \affiliation{%
Max-Planck-Institut f\"ur Physik, F\"ohringer Ring 6, 80805 M\"unchen, Germany
}%
 \affiliation{%
Center for Cosmology and Particle Physics, Department of Physics, New York University, 4 Washington Place, New York, NY 10003, USA
}%

\date{\today}

\begin{abstract}
We show that the assumption of non-zero  topological susceptibility of the vacuum in a fermion-free version of a theory,  such as gravity or QCD, suffices to conclude the following:  Once $N_f$ massless fermion flavors are added to the theory, they break the  chiral flavor symmetry dynamically,
down to a subgroup that would be anomaly-free under gauging;
 In both theories,  the pseudo-Goldstone corresponding to axial $U(1)$-symmetry becomes massive;  
In QCD as well as in gravity the massless fermions are eliminated from the low energy spectrum of the theory.  
All the above conclusions are reached without making an assumption about confinement.   
  Some key methods of our approach are: 
Reformulation of topological susceptibility in the language of a $3$-form gauge theory;
Utilization of gravity in the role of a spectator interaction  for the chiral anomaly-matching in QCD;  Gauging chiral symmetries  and matching their anomalies using the spectator Green-Schwarz type axions. 
 Our observations suggest that breaking of chiral symmetries 
 in QCD and in gravity can be described in unified topological language, 
   and seemingly-disconnected phenomena, such as, the generation of $\eta'$-mass
 in QCD  and breaking of global chiral  
symmetry by gravity may share a secret  analogy. 
The described phenomenon may shed a new light  - via contribution of micro black holes into the gravitational topological susceptibility of the vacuum -  on incompatibility between black holes and global symmetries. It appears that explicit breaking is not the sole possibility, and like QCD,  gravity may break global symmetries dynamically. 
As an useful byproduct,  matching of gravitational anomalies provides a selection tool for compositeness, eliminating possibility of massless composite fermions where standard gauge anomaly matching would allow for their existence.

\end{abstract}

%\pacs{(old) 14.60.Pq,13.15.+g,04.60.-m,11.30.Rd}

\maketitle

\section{Introduction}
An important role of topological susceptibility of the vacuum in defining the infrared properties of QCD has 
been long appreciated.  A well-known example is the generation of the mass of 
the $\eta'$-meson via Witten-Veneziano mechanism \cite{WV}, 
 in the framework of QCD with 't Hooft's large number of colors
\cite{planar}.  
This approach, in difference with the original solution by 't Hooft \cite{tHooft1976, determinant},  does not explicitly rely on instantons, but on the 
topological susceptibility of the vacuum.   
Also, a direct connection between this entity and 
appearance of Goldstones with simultaneous solution to $U(1)$ problem 
in QCD,  has been suggested by Veneziano in \cite{veneziano}.  
 
Yet, the power of vacuum topological susceptibility has not been fully exploited for understanding the  
vacuum structure and the dynamics of chiral symmetry breaking.

First, to our knowledge, the question whether the assumption about 
vacuum topological susceptibility suffice for proving the further breaking of $SU(N_f)_L\times SU(N_f)_R$ chiral symmetry in QCD with $N_f$ massless  quark flavor, 
is still open. 
 
Moreover, the role of a gravitational version of the topological susceptibility of the vacuum  in generation of the mass gap and chiral symmetry breaking by fermion condensate was explored only relatively recently \cite{gia3form, DJP, DFF,DFneu}.  
 
  In this note we would like to expand on previous results and show that  topological susceptibility of the vacuum   gives a powerful tool both for establishing the fact 
 and predicting the pattern of dynamical chiral symmetry breaking  in generic class of theories.  
 
 Some steps 
 were already taken in this direction.  It was shown \cite{gia3form} that in QCD the 
vacuum topological susceptibility allows to understand and re-formulate  the mass generation -
 both for $\eta'$-meson as well as for axion  - {\it entirely}  in the language of 
 a Higgs phase of a $3$-form gauge theory.   
  Moreover, it was 
 also shown there that the same connection persists in gravity:  The existence of 
 vacuum topological susceptibility in pure gravity, would imply the generation of mass gap with the participation of all the sectors (such as, chiral fermions or elementary axions)  
of the theory exhibiting a chiral gravitational anomaly. That is, a collective pseudo-scalar degree of freedom - transforming  under  the  axial symmetry anomalous with respect to gravity  - would become  massive.

 It was further shown \cite{DJP} that the connection between the 
 existence of topological Chern-Pontryagin density and generation of mass gap due to  
 anomalous current is generic  and can be formulated entirely in topological terms:  In any theory 
 with vacuum topological susceptibility and an anomalous current, one can predict a generation of mass of a pseudo-scalar meson corresponding to a would-be Goldstone of the anomalous current.

  The sectors most sensitive to the gravitational topological susceptibility would be 
axion and neutrinos.  As shown in \cite{gia3form, DFF}, the existence of such topological susceptibility in the absence of massless fermions,  
would directly contribute into the mass of the axion, and thus, would jeopardize the Peccei-Quinn solution \cite{PQ} to the strong CP problem.
  As a way out \cite{gia3form, DFF}, in the presence of fermions with zero or small bare masses (e.g., a single neutrino 
 species would suffice) the axion mass becomes protected:  Such fermions contribute into the chiral gravitational anomaly 
 and nullify the gravitational topological susceptibility of the vacuum. Simultaneously, the fermion sector must deliver a pseudo-scalar 
  would-be Goldstone boson  corresponding to the anomalous current. 
  The mass of this boson is generated in a way very similar to the generation of $\eta'$ mass in QCD.

  The connection between the vacuum topological susceptibility in gravity and generation of mass gap for light fermions may find an interesting phenomenological application for neutrino masses \cite{DFneu} and neutrino-axion physics\cite{DFF}.     
  In this light, settling the question of gravitational vacuum topological susceptibility
 and its role in chiral symmetry breaking is of direct important for phenomenology.  
   
   Summarizing  where we stand so far:   The above studies showed that  if the fermion-free version of a theory 
  (e.g.,  QCD or gravity ) admits a non-zero value of topological susceptibility, then after introduction of fermions, they condense and break the anomalous  
 $U(1)$ chiral symmetry spontaneously. In the same time 
 a would-be Goldstone boson acquires a mass due to a $3$-form Higgs effect.

  On the other hand, to the best of our knowledge,  the necessity of breaking - solely due to vacuum topological  susceptibility - 
    of chiral symmetry beyond the anomalous $U(1)_A$ has not been established firmly, neither in gauge theories nor in gravity.        
 
In this note we shall attempt to fill up this gap and show  that from the  assumption of non-zero  topological susceptibility of the vacuum in a fermion-free version of a theory,  such as QCD or gravity, follows the following: 
\begin{itemize}
  \item Once $N_f$ massless fermion flavors are added to the theory, they form a condensate breaking the $U(N_f)$ chiral flavor symmetry  
  dynamically,  down to one of its maximal anomaly-free subgroups, i.e., a maximal subgroup  
 that, if gauged, would be anomaly-free.  
   
  \item  In both theories,  the pseudo-Goldstone corresponding to axial $U(1)$-symmetry becomes massive. This generation of mass can be understood as a $3$-form Higgs effect.   
 
  \item Both in gravity and in QCD the massless fermions are eliminated from the low energy spectrum. 
  
  \item  All the above phenomena are independent  of assumption about the confinement.

\end{itemize}

 We shall speculate that the dynamical breaking of chiral symmetry due to topological susceptibility of the vacuum may shed a new light on incompatibility between black holes and global symmetries.  The link could be provided by a contribution into the gravitational topological susceptibility of the vacuum from the micro black holes. That is, black holes could break global chiral symmetries 
 dynamically via such a contribution.
  
  This adds a new flavor to the relation between black holes and global symmetries. 
   The standard assumption is an explicit breaking of such symmetries by some high-dimensional operators.  What we observe is that, similarly to QCD, the breaking 
   of global chiral symmetry in gravity may be {\it dynamical}.  Of course, 
  the latter does not exclude the possibility of explicit breaking. In fact, as we shall see, for the axial $U(1)$ both types of breaking (explicit and spontaneous-dynamical) originate from the gravitational topological susceptibility of the vacuum.  
   The $3$-form language nicely clarifies the coexistence of these two types of breaking:  The generation of mass for a would-be Goldstone boson
from explicit breaking  by gravitational anomaly,  can be viewed -
in alternative language - as the mass coming from the $3$-form Higgs effect.    
  \\

  Let us briefly mention some of  the new approaches that we 
 use in our treatment. 

  In our  analysis, we shall use the formulation \cite{gia3form}, which describes   
 phenomena of strong CP-violation and generation of the mass of 
 $\eta'$-meson  (or axion) in the language of  
a $3$-form gauge field.   
   The convenience 
of this language is that it allows 
to understand the vacuum structure of QCD with and without massless quarks as the two phases  - Higgs and  Coulomb - of a  $3$-form gauge theory.

 We must note that the possibility of modeling the $\eta'$ mass in QCD by a massive $3$-form, 
 has been recognized for some time, in particular, by Aurilia, Takahashi  and Townsend \cite{aurilia}.   Similar effects have been noticed by several authors 
in different contexts. For example,  Gabadadze and Shifman discussed  the screening  of $3$-form by axion, while analyzing the structure of axion domain walls \cite{walls}.    
 
  The analysis in \cite{gia3form} is important for us in two respects.  
 First, it shows that $3$-form Higgs effect accounts for the entire QCD contribution to the $\eta'$ and axion masses, including the one coming from 't Hooft's instantons, as well as, for an additional contribution in case of a non-zero 
gravitational topological susceptibility of the vacuum. Secondly, it fully accounts for the CP-invariance of QCD (or gravitational) vacuum.  
  So it appears that the $3$-form Higgs language is not just an useful analogy for modeling the $\eta'$ (or axion) mass generation in QCD (or a similar effect in gravity), but 
  rather, it represents a remarkably precise effective description of this phenomenon.  
  
  We shall elaborate on this issue below.     
The presentation shall be
self-sufficient, but for a more detailed discussion of this    
formulation the reader is referred  to \cite{gia3form}. \\ 

The other two important ingredient of our analysis are the following. \\
 
First,  we shall heavily rely on  't Hooft's anomaly matching \cite{anomalymatching}, but shall use as spectators 
not only fermions, but also Green-Schwarz type \cite{GS}  axions.  This allows us 
to weakly gauge and monitor the anomaly matching even with respect to the symmetries 
that are anomalous and are explicitly broken by instantons \cite{determinant,
tHooft1976} or other effects.   

This is important, since for example, in his original 
breakthrough paper on anomaly matching \cite{anomalymatching}, 
't Hooft did not impose anomaly matching for the axial $U(1)_A$-symmetry, because it is explicitly
broken by instantons.  However, as we show below,  for gauged 
$U(1)_A$ for which anomalies are cancelled by a spectator Green-Schwarz axion, 
the  't Hooft determinant term generated by the instantons \cite{determinant} 
 is manifestly gauge invariant under $U(1)_A$. Therefore,  anomalies under this symmetry must be monitored. This  monitoring provides us with some crucial information, for controlling the breaking of the non-abelian $SU(N_f)$ part of the chiral symmetry. \\

 Another new tool that we employ in  QCD, is that we use {\it gravity}  as a spectator interaction and impose matching of the gravitational chiral anomalies.  
 The extra power provided by this method relies on the fact that, although composite fermions can be color-singlets, they cannot be ``gravity-singlets". By general covariance, there exist no particles decoupled from gravity. 
 
 This gives an additional consistency relation and eventually enables us to deduce 
 the absence of massless fermions, both elementary and composite, and  
 a complete breaking of any part of chiral symmetry that would exhibit anomalies  
under gauging. \\

We must note that the employment of gravity in the role of  a spectator interaction for the anomaly-matching, is a byproduct that can be used in composite model building 
 irrespectively of the issue of  vacuum topological susceptibility.  Even if one ignores the latter and uses the confinement or a compositeness as a starting assumption, the gravitational anomaly can severely restrict the composite spectrum, e.g., by eliminating massless composites  in cases where standard gauge anomaly-matching would allow their existence.

\section{Assumption and  reasoning}

 Let us first formulate our assumptions.  
We consider a generic theory (gauge or gravity) which satisfies the following conditions.  In the version of the theory that includes no massless fermions, there exists 
a non-vanishing  topological susceptibility of the vacuum with respect 
to some $3$-form $C$.   As usual,  the  gauge-invariant zero-form field strength is $E \equiv ^*dC$, where the star stands for Hodge-dual.  That is, we assume that there exist the following correlator,  
\begin{equation}\label{Etop}
\langle E, E\rangle_{p\to 0}\, = \, const \neq 0  \, .
\end{equation}
where $p$ is momentum.  

 Next, assume that after introduction of $N_f$-flavors of some light fermions  there exists a chiral $U(1)_A$-current with the anomalous divergence 
 \begin{equation}
      \partial^{\mu} J_{\mu}^{(A)} \, =  \, E \, , 
    \label{divJ} 
    \end{equation}
where for compactness we absorbed the unknown anomaly coefficient in the normalization.  Our conclusions are independent of its value as long as it is non-zero. \\
  
We then wish to prove that in such a theory fermions condense and  spontaneously break chiral symmetry.  The 
would-be Goldstone of the anomalous current (call it $\eta'$) is becoming massive. 
It is eaten-up by  the three-form $C$ and becomes massive via an effect closely analogous to  a three-form version of the Higgs 
(or St\"uckelberg) phenomenon \cite{gia3form}. 
Moreover,  the fermions are eliminated from the low energy theory. 
The above  picture of chiral symmetry breaking is independent from underlying structure of the theory.   Therefore, it is equally applicable to gauge theories as well as to gravity.  

In addition we wish to show that in QCD, the $U(N_f)_L \times U(N_f)_R$ chiral symmetry is broken  down to its subgroup that under gauging is anomaly-free.  
%vector-like subgroup $U(N_f)_{L+R}\times Z_2$, resulting into  
%$N_f^2 -1$ exactly massless Goldstone bosons and one massive one. 

In gravity,  the same holds.  
With  $N$ massless fermions, the flavor symmetry is $U(N)$, which is spontaneously broken down to an anomaly-free subgroup.  
The natural candidate for such a subgroup is $SO(N)$. 
The symmetry breaking
$U(N) \rightarrow SO(N)$ results in $N(N+1)/2$ would-be Goldstones.   Out of these, one is getting its mass directly from nullifying the topological susceptibility. 
  The remaining $N(N+1)/2 -1$ Goldstones could potentially get masses in case there exist operators that break 
 chiral flavor symmetry explicitly.   Such  
explicit-breaking sources  are fully compatible with out treatment, but
are not required for the consistency of reasoning.  Therefore, $N(N+1)/2 - 1$  Goldstones could  potentially stay
massless.  \\

  Although, we shall try to formulate our arguments in maximally generic way, 
 at first,  we shall keep making contact with QCD and later generalize to gravity. \\  
 
  It shall become clear that the arguments leading to the chiral symmetry breaking and the elimination of massless fermions should equally apply to any theory that satisfies the conditions (\ref{Etop}) and (\ref{divJ}), regardless of the assumption about the confinement.  However, the patters of symmetry breaking 
 can vary from theory to theory.  \\

   We start by more detailed formulation of the initial assumption.    
   Consider a theory that contains a gauge three-form $C$, either elementary or composite, and no massless fermions present. 
  A good example is QCD of pure glue, i.e., no quarks included. 
  We shall not specify the number of colors, and 
  allow ourselves to take it large, when needed. 
   We assume that the topological susceptibility of the vacuum is non-zero, 
   (\ref{Etop}). 
      
  For QCD the condition (\ref{Etop}) takes the following form, 
 \begin{eqnarray}
  % \begin{equation}
   \label{FFcorr}
\langle F\tilde{F},F\tilde{F}\rangle_{p\to 0} \equiv \nonumber\\
\equiv \lim\limits_{p \to 0}\int d^4 x e^{ipx} \langle T [F\tilde{F}(x)F\tilde{F}(0)]\rangle =\mathrm{const}\neq 0\, , 
\end{eqnarray}
%\end{equation}
where $F$ is the usual QCD field strength two-form, $\tilde{F} \equiv ^*F$. \\  

   One important thing that a non-zero topological susceptibility of the vacuum
  (\ref{FFcorr}) tells us, is that  the theory contains vacua that break CP-symmetry.
   These vacua are usually labeled by a CP-violating parameter $\theta$. 
 The physical observability of this parameter is equivalent to (\ref{FFcorr}). 
 Correspondingly, when the correlator in (\ref{FFcorr}) vanishes, 
 CP-violation becomes unobservable and one can say that $\theta$-parameter  becomes unphysical.

   Throughout our discussion we shall heavily rely on matching the 
 descriptions - of certain physical phenomena  - in the languages of  high-energy and low-energy theories.  
   In particular, these two descriptions must agree on the fact that  in the absence of massless quarks the CP-violation is {\it physical} and that  it becomes {\it unphysical}  once the  massless quarks are introduced.   This matching of the two descriptions will  lead us to the conclusion of necessity of chiral symmetry breaking by fermion condensate  and  of elimination of massless fermions from the low energy theory.   
   We shall now discuss this in more details.

\subsection{Alice and Bob} 

    Let us introduce the two observers:  A high energy observer Alice and a low energy 
  observer Bob.  These two physicists must both describe the same physics, but from two different perspectives. 
  
  In particular, they both agree that in QCD without quarks there exist CP-violating 
  vacua.   For Alice, this is accounted by the introduction of the $\theta$-term in 
  her Lagrangian,
  \begin{equation}
  L_{Alice} \, = \,  \theta F\tilde{F} \,  -  \, FF \,, 
 \label{AliceT}
 \end{equation}
 where the second term is the usual QCD field strength, with irrelevant constants absorbed in normalization.     
In the absence of quarks, the $\theta$-term is physical and can be measured. 
\footnote{For example, this can be accomplished by introduction of  some heavy external sources,  (e.g., a heavy pseudo-scalar) coupled to $F\tilde{F}$, and by the measurement of  CP-violating effects in their interactions.  The strength of these effects will be controlled by $\theta$. The precise methodology is unimportant for our discussion.}    
 \\

 Let us  explain how the same CP-violating vacua are described by Bob.   

For this, let us rewrite the theory in terms of the  Chern-Simons three-form  $C$ and its field strength  $E \equiv ^*dC$, 
\begin{equation}
C\equiv A\wedge dA + {2 \over 3} A \wedge A \wedge A,~~~ 
E \equiv F\tilde{F}=^*\mathrm{d}C.\label{eqE}
\end{equation}
Here, $A$ is the gluon gauge field one-form. 
Under the QCD gauge transformation $C$ shifts by an exterior derivative of a two-form,  
\begin{equation}   
C \rightarrow C +  d\omega \,, 
\label{gauge}
\end{equation}
where the explicit form of $\omega$ is irrelevant and shall not  be displayed here.    
When expressed in terms of the field $C$, the non-vanishing correlator (\ref{FFcorr}) implies that the propagator of $3$-form $C$ has a pole at $p^2\, =\, 0$,
\begin{equation}\label{CCcorr}
\langle C, C\rangle_{p\to 0}\, \propto \, \frac{1}{p^2} \, .
\end{equation}
As noticed by L\"uscher  \cite{massless3form},  this fact tells us that $C$ is a fully legitimate {\it massless}  $3$-form gauge field, with all the usual properties. 
If we couple $C$ to some external probe
sources, then at large distances $C$ will mediate interactions between them exactly  the way an elementary $3$-form does. 
For example, if we source $C$ by a $3$-form current of some probe membrane, it will induce a distance-independent static  force between the two parallel  membranes
(see, \cite{gia3form}). This force is very similar to static  Coulomb force between 
the two point-like massive charges in $1+1$-dimensional Schwinger model. The similarity is not accidental, since the massless one-form field in $1+1$  does not propagate any degree of freedom.  The latter property is fully shared by the  massless $3$-form in $3+1$-dimensions.  

  Because of this, the existence of a massless  $3$-form  in effective theory 
is in no conflict with usual intuition that a theory of a pure glue has a mass gap. 
Although a massless $3$-form does not propagate any physical degree of freedom,   nevertheless, it can create a static long-range electric-type field,
as discussed above.  As we shall see,  this static field is the key for describing CP-violation in low energy theory.

 The above knowledge combined with the invariance of the theory under the gauge shift  (\ref{gauge}) leads to the effective Lagrangian describing the topological structure of the QCD vacuum in form of a gauge theory containing solely the massless three-form $C$ \cite{gia3form}. 
In other words, at zero momentum all the massive states decouple and the vacuum 
structure of the theory is fully determined by  its dependence on $C$.

 Since we are interested in the vacuum structure, i.e.,  the  zero momentum limit of the theory, we shall work in effective low energy theory below the QCD scale, with all the massive excitations being integrated out. 
 
  We shall relax the assumption about the mass-gap shortly, but it is useful to 
  first go thought the argument under it.  
 Since we are working in theory with a mass-gap, there are no propagating  massless  
degrees of freedom in this theory. 
Then, by gauge invariance, the effective Lagrangian includes 
the following terms:  1) A function algebraic in $E$; and 2) the terms that depend on derivatives of $E$. 
Since no massless excitations have been integrated out, 
all the higher derivative terms must vanish in zero momentum limit.

  Let the algebraic function be ${\mathcal K}(E)$. Then,  the effective Lagrangian has the form, 
\begin{equation}
\label{effQCD}
L_{Bob}  \, = \,  {\mathcal K}(E) + \, ({\rm terms~depending~on~}  \partial_{\mu} E ) \,. 
\end{equation}
 %For simplicity, we have written the equation in units of QCD scale.  
 We shall now look for the vacuum solutions of the equations of motion
 obtained by taking variation with respect to $C$.  These 
 are solutions  with constant $E$. Because of this, the contribution from the derivative terms is automatically zero and the 
 equation of motion reduces to,  
 \begin{equation}\label{eqEeff}
 \partial_{\mu} \left ( {\partial {\mathcal K}(E) \over \partial E} \right ) \, = \, 0,
\end{equation}
  This equation is satisfied for an arbitrary constant value of  $E$  
  (except the values for which derivative of ${\mathcal K}$-function may be singular). 
  From now on we shall work in units of the scale $\Lambda$ that sets the 
  relation $F\tilde{F}  =  E\Lambda^2$, where $E = ^*dC$ is written for
  $C$ with canonical normalization.  The value of this scale is determined 
  by the vacuum topological susceptibility and its precise value is unimportant for us.    
    In units of this scale,  we can parameterize the constant vacuum expectation value  (VEV) of the electric field that solves (\ref{eqEeff}) as 
 \begin{equation}
      E = \theta\,.     
  \label{Tvac}
  \end{equation} 
  Thus, the vacua of the theory are labeled by an integration constant $\theta$ and 
  correspond to vacua with different values of the static ``electric field"  $E$
  \footnote{The term ``electric field" is used in the analogy with electrodynamics, 
  with $C$ playing the role of the electromagnetic vector potential.  For example, the constant value of  $E \equiv  \epsilon^{\alpha\beta\mu\nu} \partial_{\alpha } C_{\beta\mu\nu} $ corresponds to a choice $C_{\mu\nu\alpha} = \epsilon_{\mu\nu\alpha\beta} x^{\beta}$. }

 Notice that the vacua with different values of $\theta$  are physically distinct, since they differ by the amount of CP violation:  The electric field $E$ is CP-odd.
   We may try to impose the condition that the initial function ${\mathcal K}(E)$ is CP-invariant, meaning that it only contains even powers of $E$,  ${\mathcal K}(E) = {1 \over 2} E^2 + ...$. 
   Nevertheless, re-expanding the theory  around the vacuum $E=\theta$, we end up with a linear 
CP-violating term in the Lagrangian,  
 \begin{equation}\label{EXP}
  L_{Bob} \, = \, \theta E  +  {1 \over 2} E^2 + ... \,.
\end{equation} 
For the purpose of our discussion we can limit ourselves with infinitesimal  values 
of $\theta$  (and correspondingly  $E$).  \\

The direct connection between the existence of non-zero topological susceptibility (\ref{FFcorr}) and a static electric field $E \neq 0$,
removes any need for the  assumption of the mass gap, which we shall now relax.  

 Indeed, the existence of a constant electric field $E$ means that in the theory 
 (in this case pure glue)  there exist 
  no {\it mobile charges}  that source the $3$-form $C$. If such charges 
 were to  exist, 
  they would give the mass to $C$ and would correspondingly screen the electric field $E$, in a way very similar to what happens to an ordinary Maxwell electric field 
  in a superconductor  and/or in the Higgs phase. 
  
  This can be understood directly at the level of the effective Lagrangian for $C$ and is very important.  Naively, one could say that if we do not demand a mass gap, then 
  the effective  Lagrangian for $C$ could contain ``non-local" 
 terms resulting from integrating-out some massless degrees of freedom, 
 which could potentially ruin the static electric field solutions.  
 However, the gauge-invariant ``non-local" terms that could do such a job
 (e.g., the terms of the type $C \Pi C$, where 
  $\Pi$ is a transverse projector) can be explicitly shown to come  
from integrating-out the mobile sources of $C$, such as,  massless axions or massless membranes (see \cite{gia3form}).  But,  their existence would be equivalent of having zero topological susceptibility of the vacuum in theory of pure glue, which would contradict to our starting assumption.      
  
  In other words, the screening of $E$ would imply that the pole at $p^2=0$ is removed, or equivalently,  the vacuum topological susceptibility (\ref{FFcorr})  would vanish.  
 This argument shows that the existence of the constant vacuum electric field 
 $E=\theta$ is a direct consequence of non-zero vacuum topological  
susceptibility  (\ref{FFcorr}) and does not require any extra assumption about a mass gap in the theory.  Conversely,  introduction of mobile charges that screen 
$E$, would render $C$ massive,  and thus, would make the topological susceptibility of vacuum 
zero.  As we shall see later, this is exactly what happens once we introduce massless fermions in the theory.  However, for the time being, we restrict ourselves by pure glue.   

    Since in the theory there exist no dynamical sources 
for $C$,  there are no transitions between the vacua with different  $\theta$.  Hence,  the theory splits into infinity of super-selection sectors  labeled by the angle $\theta$.   These vacua are nothing but the famous  $\theta$-vacua of QCD \cite{tHooft1976, 
thetavacua}, which Bob has  
``rediscovered" in the language of a $3$-form gauge theory.  These 
are exactly the same vacua that Alice labels by the $\theta$-term in (\ref{AliceT}). 

 We thus arrive to the following matching of the descriptions of the two
 physicists:  Alice's $\theta$-vacua are understood by Bob as the vacua with different 
 values of the static CP-violating  electric field $E$, 
 \begin{equation} 
 \left (\theta-{\rm vacua} \right )_{{\rm Alice}}\,  =  \, \left (E-{\rm vacua} \right )_{{\rm Bob}} \, . 
   \label{AB}
   \end{equation} 
   That is,  what for Alice serves as a parameter of the theory, for Bob is a label of 
the vacuum.  When Alice changes the theory by choosing a different value of  $\theta$,   Bob is making a corresponding choice by selecting the vacuum with a different value of the electric field $E$.  \\

[Here one could be tempted to enter into a semantics discussion and ask: 
Which term is more appropriate for the use with respect to $\theta$, 
a ``parameter" or an ``integration constant"?  Usually, the difference 
between the two notions is that the former labels a theory, whereas the latter
 labels a particular solution (in this case a static electric field) in  a given theory. 
 However, when the choices obey a super-selection rule, the two notions become equivalent.  For example, since there exist no charges that could source $C$ in theory of pure glue,  no transitions occur  between the sectors with different 
 values of $E$. The Hilbert spaces constructed on top of these vacua are 
 orthogonal.  So Bob could equally well call  $E$ a parameter, rather than a ``solution".   The two languages describe the same physics,
 as they should, and matching the two descriptions gives us an useful
 tool.] \\

    Notice, the vacua have different energies, but due to absence of sources 
    there are no transitions among them.   The lowest energy vacuum is the CP conserving one.  Alice explains this using Vafa-Witten theorem \cite{VafaWitten}. 
    For Bob, the same conclusion is reached because the lowest energy state 
    is the one with the vanishing electric field, $E = 0$.     
    
Notice, as explained in \cite{gia3form}, the requirement that $\theta$ be defined modulo $2\pi$ translates as a constraint on the ${\mathcal K}$-function that 
its derivative must be inverse of a periodic function.  Since we are working 
with small values of $E$, this constraint  is not important for our discussion. 
\\
  \subsection{Introducing Massless Quarks} 
    
  Let us now introduce in the theory some massless quark flavors. It is useful to introduce them in form of  left-handed Weyl fermions 
 in fundamental $\psi_{i}$ and anti-fundamental $\bar{\psi}^{\bar{j}}$ representations of the color group.  Here the bar does not stand for Dirac conjugation, but for    
 distinguishing  the fundamental and anti-fundamental representations of the color group.  $i,j = 1,2,...N_f$ and 
 $\bar{i},\bar{j} = 1,2, ...N_f$ are the flavor indexes. The color index is not written explicitly. 
% Some times,  for the compactness of the notations, we shall pair them up into 
% Dirac fermion flavor $\Psi_{(j)}$. 
 In the absence of the fermion masses the theory exhibits a  
 $U(N_f)_L\times U(N_f)_R$ flavor symmetry, where $U(N_f)_L$ and  $U(N_f)_R$
 act on $\psi_{i}$ and $\bar{\psi}^{\bar{j}}$ independently.  The fermions thus form 
 a bi-fundamental representation $(\bar{N}_f, N_f)$ of the flavor group.
 It is useful to write the flavor group as $SU(N_f)_L\times SU(N_f)_R \times U(1)_V\times U(1)_A$.  Where, $U(1)_V$ and $U(1)_{A}$ are flavor-blind vector and axial symmetries that act on the fermions as 
 \begin{equation}
    \psi_{i} \rightarrow {\rm e}^{i\alpha}  \psi_{i}, ~~~ \bar{\psi}^{\bar{j}}  \rightarrow {\rm e}^{-i\alpha} \bar{\psi}^{\bar{j}} 
   \label{vector}  
  \end{equation}
  and 
    \begin{equation}
    \psi_{i} \rightarrow {\rm e}^{i\alpha}  \psi_{i}, ~~~ \bar{\psi}^{\bar{j}}  \rightarrow {\rm e}^{ i\alpha} \bar{\psi}^{\bar{j}} 
   \label{axial}  
  \end{equation}
 respectively.  Here $\alpha$,  in each case,   is a transformation parameter.

 We now wish to show that the fermions form a condensate that 
 breaks  
  $U(N_f)_L\times U(N_f)_R$ chiral symmetry spontaneously
   at least down to its anomaly-free subgroup,  i.e., a subgroup that would be anomaly free if the entire  $U(N_f)_L\times U(N_f)_R$-symmetry 
were to be gauged. \\

   We shall organize the argument in four steps: \\
   
  \begin{itemize}
 
  \item First, we shall make sure that the introduction of quarks makes the vacuum topological susceptibility zero. Both for Alice and Bob this implies 
  that vacuum becomes CP-conserving.  
  
  \item  Next, following \cite{gia3form} we show that in Bob's description
  this means that $3$-form gauge theory enters the Higgs phase, i.e., 
  $C$ acquires a non-zero mass.  By gauge-invariance (\ref{gauge})  this is only possible if $C$ ``eats" a  pseudo-scalar degree of freedom, which we denote by $\eta'$. We determine that the only degree of freedom that  matches the required properties is the phase of the quark-condensate.  Hence, quark condensate must
break $U(1)_A$-symmetry spontaneously.  
  This part  of the argument reinforces the results already presented in \cite{gia3form, DJP, DFF, DFneu}.  It is also in  agreement  with the conclusion of \cite{veneziano}, reached in a different approach.  
  
   \item   Next,  we do two things. First, we introduce gravity as a spectator interaction. 
   Secondly, we gauge the axial symmetry weakly and cancel the gauge-gauge and 
   mixed gauge-gravity anomalies by a spectator Green-Schwarz axions. 
   Then, by matching anomalies in theories of Alice and Bob  we conclude that 
  there cannot exist any massless fermions at low energies. 
  
  \item  Finally, by weakly gauging the different subgroups of the flavor group and 
  applying 't Hooft's anomaly matching conditions with spectator fermions, we
  deduce  that  the quark condensate must break the 
   $U(N_f)_L \otimes U(N_f)_R$-symmetry down to its anomaly-free subgroup.

   \end{itemize}

 \subsection{Necessity of fermion condensate and mass-generation for $\eta'$} 
   
  Both  Alice and Bob must agree on the fact that after introduction of the massless quarks the topological susceptibility must vanish, or equivalently, 
 the parameter $\theta$ must become unphysical.
  This is an effect that can be established experimentally: $E$ is a CP-odd quantity, which can be measured.  Thus, unambiguously, after introduction of massless fermions  both Alice and Bob must measure zero CP-violation in their theories.

  However, they explain this fact differently. \\
   
 Alice knows that  $\theta$ is unphysical because it can be rotated away by the chiral transformation (\ref{axial}).  Due to Adler-Bell-Jackiw chiral anomaly 
 \cite{ABJ}
 the Lagrangian shifts under this transformation as 
 \begin{equation} 
  \delta L_{Alice} \, = \, \alpha F\tilde{F} \, =  \alpha E \, ,
   \label{shiftC}
 \end{equation}   
    and any initial $\theta$-term can be eliminated by the suitable choice of 
  $\alpha$.   Here and below, we absorb the anomaly coefficient in definition 
  of $F\tilde{F}$. 
  
   One of the consequences of this freedom is that CP is unbroken.  
   Thus, the topological susceptibility must vanish.    
   This means 
 that the pole in $3$-form correlator must be moved away from  $p^2=0$, 
  \begin{equation}\label{eq:CGap}
\langle C, C\rangle_{p\to 0}=\frac{1}{p^2 - m^2}\;.
\end{equation} \\

 The equation (\ref{eq:CGap}) explains how the same phenomenon is accounted 
 by Bob.  
 For Bob 
 the physicality of $\theta$ was equivalent to the fact that vacuum could support a constant electric field 
 $E$, which breaks CP-symmetry.  This is only possible if  the low energy theory contains  a massless three-form $C$. 
 The elimination of  $\theta$ - that Alice understands as 
 a freedom of changing the fermion phases by a chiral rotation (\ref{axial}) - 
  for Bob means that the  $3$-form $C$ became {\it massive}. 
  And indeed, this is fully confirmed by the equation (\ref{eq:CGap}).  
 Thus, in order to match  Alice's story, Bob needs to explain why $C$ has acquired a non-zero mass after quarks were introduced. 
  How can a $3$-form gauge field $C$ acquire a mass?  
 
 The massive $3$-form, unlike the massless one, 
 does propagate one (pseudo-scalar) degree of freedom.  The key point is that the generation of the mass  gap must happen 
 with the full respect of the gauge redundancy (\ref{gauge}), i.e.,  the $3$-form must acquire the longitudinal  pseudo-scalar degree of freedom as a result of a Higgs-like effect. 
 Let us denote this - yet to be identified - pseudo-scalar degree of freedom by $\eta'$.  Then, 
  up to irrelevant higher order terms, the 
 gauge invariant Lagrangian describing the mass-generation of the three-form is uniquely fixed as \cite{gia3form}, 
\begin{equation}\label{effectiveQCDeta}
L \, = \, {1 \over 2}  E^2 \, - \, {\eta' \over f_{\eta}}  \, E \, + \,  {1 \over 2} \partial_{\mu}{\eta'} \partial^{\mu} {\eta'},
\end{equation}
where as before we work in units of the scale $\Lambda$. 
$f_{\eta}$ is a parameter that encodes information about the origin of 
$\eta'$, which we still need to discover. 
Also notice: the Higher order derivative terms cannot modify the value of the mass gap, since they all vanish at zero momentum. 
 The equations of motions for $C$ and $\eta'$ are  
\begin{equation}\label{eqEeff}
\partial_{\mu} \left( E -  {\eta' \over  f_{\eta}} \right)=0,
\end{equation}
 and 
\begin{equation}\label{boxeta1}
\Box \, \eta' =  \, - \,  {1 \over f_{\eta} } E \,, 
\end{equation}
respectively. 
Solving the equation of motion for $C$,
we obtain the following expression for the electric field
\begin{equation}\label{solEeff}
E \, = \, \left({\eta' \over f_{\eta}}-\theta\right),
\end{equation}
where, just as in the fermion-free case (\ref{Tvac}),  the parameter $\theta$ is an integration constant. The crucial novelty that  coupling with $\eta'$ introduces is 
that $\theta$ is {\it unphysical}: 
For arbitrary choice of $\theta$, the vacuum expectation value of the electric field $E$ vanishes and simultaneously the mass gap is generated. 
 
Indeed, inserting (\ref{solEeff}) in (\ref{boxeta1}) we get the following effective equation of motion for $\eta'$,  
\begin{equation}\label{boxeta}
\Box\, \eta' +  {1 \over f_{\eta} }\left({\eta'  \over f_{\eta}} -\theta \right) = 0. 
\end{equation}
Here we immediately observe: 

 1) The fields $\eta'$ and $C$ combine and form a single propagating massive pseudo-scalar field of mass $m_{\eta'}^2 = 1/ f_{\eta}^2$; 
 
  and  
 
 2) The vacuum expectation value of this field, ${\eta' \over f_{\eta}} =\theta $, is exactly such that it forces $E = 0$, as it is clear from (\ref{solEeff}).  Thus, irrespective of the choice  
of the integration constant $\theta$, Bob detects zero CP-violation, since 
the VEV of the CP-odd electric field $E$ is zero.  Notice, this is 
a manifestation in Bob's language of Vafa-Witten theorem \cite{VafaWitten}, which implies that once $\theta$ is promoted into a dynamical field, it relaxes to CP-conserving minimum.   
 \\

 What we are witnessing is the analog of the Higgs phase for the $3$-form $C$. 
 As it is clear from the second term  in (\ref{effectiveQCDeta}) the $3$-form 
 $C$ is sourced by the topological $3$-form current 
 $d\eta'$. Thus,  the pseudo-scalar $\eta'$ plays the role of the mobile charge  that 
 screens the electric field and gives mass to $C$.   \\

    {\it To summarize:  So far - using the gauge-invariance (\ref{gauge})
   and the matching of the descriptions of Alice and Bob - we have established that  the  
 introduction of fermions is accompanied by the appearance of a pseudo-scalar $\eta'$, which renders the $3$-form $C$ massive.   } \\
   
    We now need to identify the origin of $\eta'$.
   The first step is to understand that $\eta'$ is a phase excitation of 
   some order parameter formed out of fermions, which transforms 
   non-trivially under $U(1)_A$.  That is, $\eta'$ is a would-be Goldstone boson 
   of spontaneously broken chiral  $U(1)_A$-symmetry.

 The above understanding can again be achieved by matching the descriptions of Alice and Bob.

 Alice, being an UV-observer, has no information about the existence 
 of $\eta'$.  But, she knows that $\theta$ is unphysical, because she
  can eliminate it  by arranging a shift of the Lagrangian (\ref{shiftC}) 
  by performing a chiral 
 $U(1)_A$-transformation of fermions (\ref{axial}).  So, by this transformation Alice can shift $\theta$ as 
     \begin{equation}
   \theta \rightarrow  \theta  \, + \, \alpha \,.
 \label{shiftTA}
\end{equation}

Since the CP-invariance is universal for both observers, the shift of $\theta$ by the chiral rotation performed by Alice,  
 in theory of Bob must be exactly matched by the shift of  $\eta'$.
 This is also clear from the equation (\ref{solEeff}) which shows that in Bob's theory 
 the vacuum is always  at   ${\eta' \over f_{\eta}}-\theta = 0$, where 
 $E$ vanishes.  Therefore, the shift of $\theta$  implies the  corresponding  shift    
 of $\eta'$,  such that the equality $E=0$ is kept intact. 
 
 In the normalization in which $\theta$ denotes a coefficient 
of $F\tilde{F}$ and the anomalous coupling of $\eta'$ is 
\begin{equation} 
{\eta' \over f_{\eta}} F\tilde{F} \, , 
\label{cefficient} 
\end{equation} 
 the shift of $\eta'$ induced by the  Alice's chiral transformation is  
      \begin{equation}
   {\eta' \over f_{\eta}} \rightarrow  {\eta' \over f_{\eta}} \, + \, \alpha \,.
 \label{shiftEta}
\end{equation}

The above is the key point and let us reiterate it.  For Alice the 
non-existence of physical CP-violation is in arbitrariness of   
 $\theta$, which she can change at will by the chiral rotation (\ref{axial}). 
 This shift  in the theory of Bob corresponds to 
 a shift of the integration constant 
 $\theta$ in (\ref{solEeff}).  The non-existence of CP-violation  for Bob means that 
 the zero value of the electric field $E$ is insensitive to this shift. 
 This in only possible if the shift  of $\theta$  (\ref{shiftTA})  is exactly compensated by the shift of $\eta'$ in (\ref{shiftEta}).  
 This  shows that $\eta'$ is a phase degree of freedom of  some  fermion composite 
 order parameter that transforms under $U(1)_A$. 
 The matching between the descriptions of Alice and Bob is summarized in table \ref{BHtable}. \\

%%%%%%%%%%%%%
 \begin{table}
  \begin{center}
 \begin{tabular}{|l | l|}
  \hline
  \textbf{Alice} & \textbf{Bob}\\ 
  \hline
  No fermions (physical $\theta$) & Coulomb phase  ($E \neq 0$) \\
  \hline
  Chiral Fermions (unphysical $\theta$)  &  Higgs phase ($\eta'$ screens $E$) \\
  \hline
   $U(1)_A$-shift: $\theta \rightarrow \theta + \alpha$   & $U(1)_A$-shift: 
    $\eta' \rightarrow \eta' + f_{\eta} \alpha $ \\
       \hline
\end{tabular}
  \end{center}
\caption{The table summarizing correspondence between the descriptions of Alice and Bob.}\label{BHtable}
\end{table}
    
Notice, the $3$-form language\cite{gia3form} that Bob is using to conclude 
the existence of massive pseudo-scalar $\eta'$,  fully  agrees with the standard language of 't Hooft about generation of this mass from instantons. 
 In 't Hooft's language, the explicit breaking is due to the instanton-generated  term in the effective Lagrangian \cite{determinant},
 \begin{equation} 
 L_{'t Hooft}  \, = \, e^{-i\theta} det(\bar{\psi}\psi)\,
 \rightarrow  \, e^{- i(\theta - {\eta' \over f_{\eta}})} |\langle det(\bar{\psi}\psi) \rangle |\,.  
  \label{detFermion} 
  \end{equation}
   After assuming that the VEV of the above order parameter is non-zero, this term generates the mass to exactly the same degree of freedom $\eta'$ as Bob sees. 
   The advantage of Bob's language is that the spontaneous breaking of 
    $U(1)_A$ is not an input assumption, but is a necessity, dictated by the 
    $3$-form Higgs effect.  In addition,  this language captures the entire 
 QCD  contribution to the mass of $\eta'$ that comes through axial anomaly, since all such sources must contribute
 through (\ref{FFcorr}) and thus must be accounted by Bob's effective 
 Lagrangian (\ref{effectiveQCDeta}).  
 
  Indeed, imagine that there exists an additional contribution, $\mu_{\eta}$, to the mass of $\eta'$,  which is not accounted by the $3$-form Higgs effect. Then, such contribution would be added to (\ref{effectiveQCDeta}) in form of an additional mass term  ${1 \over 2} \mu_{\eta}^2 \eta'^2$ and correspondingly would modify the equation (\ref{boxeta}) as
  \begin{equation}\label{boxeta1}
\Box\, \eta' +  {1 \over f_{\eta} }\left({\eta'  \over f_{\eta}} -\theta \right) \, + \, 
\mu_{\eta}^2 \eta' \, = \, 0. 
\end{equation}
The CP-odd electric field  then would have a non-zero VEV:
\begin{equation}\label{Eneq0}
E \, = \, - \, \theta {(\mu_{\eta}f_{\eta})^2 \over 1 + (\mu_{\eta} f_{\eta})^2 } \, .
\end{equation}
 This would make no sense, since it is incompatible with Alice's description, which says that massless fermions remove CP violation.  Thus,  the contribution from 't Hooft  determinant to the mass of $\eta'$ is already 
 fully included in (\ref{effectiveQCDeta}) and should  {\it not} be counted as extra. 
  
    The above shows that one should be very careful not to superimpose the two languages. This will lead 
  to double-counting! \\

%%%%%%%%

 Coming back to were we are, we have concluded that the pseudo-scalar 
 $\eta'$ that Higgses the $3$-form $C$, must come from the phase of an order 
 parameter  composed out of fermions.  However, at this point we do not know which order parameter is the right one.   
 %Our ignorance about this in encoded in  a so-far-unknown coefficient $f_{\eta}^{-1}$.
 For example, if the only non-zero order parameter were the 
 entire 't Hooft fermion determinant of $N_f$-flavors, the  $Z_{2N_f}$ subgroup 
 of $U(1)_A$-symmetry  together with $SU(N_f)_L\times SU(N_f)_R\times U(1)_V$-group  would survive unbroken.    
 %and in this case we would have $c_A = 1$. 
 In such a case  $\eta'$ would represent the over-all phase of  the entire fermion determinant.  If instead, for example, the order parameter were an universal fermion bilinear  $\langle \bar{\psi}\psi \rangle$, the flavor symmetry 
 $U(1)_A$ would be broken all the way down to  
 $SU(N_f)_{L+R} \otimes U(1)_V\times Z_2$. \\

 [As a consistency check, matching equation (\ref{boxeta1}) with the anomalous divergence  of the axial current  (\ref{divJ}) 
  we get   
   \begin{equation}
    \partial_{\mu} \eta' \, \propto  \, J_{\mu}^{(A)}\, ,  
\label{etacurrent}
\end{equation}  
  which reconfirms that $\eta'$ is a Goldstone boson of broken $U(1)_A$ axial symmetry.] \\
 
 Thus, the quark sector of the theory must deliver an order parameter 
that transforms non-trivially under $U(1)_A$.  \\

\subsection{Breaking of 
$SU(N_f)_L \otimes SU(N_f)_R$ flavor symmetry.}  

 What we need to prove now is that this order parameter must break 
 not only $U(1)_A$, but also the chiral flavor symmetry
 $SU(N_f)_L \otimes SU(N_f)_R$  at least down to its anomaly-free subgroup.    

  Introducing a notation $\Phi_{i}^{\bar{j}} \equiv \bar{\psi}^{\bar{j}} \psi_{i}$, we can construct a  Lorentz-invariant  order parameter that is invariant under 
  $SU(N_f)_L \otimes SU(N_f)_R \otimes U(1)_V$ and transforms nontrivially 
  under $U(1)_A$. This order parameter has the form of 't Hooft determinant: 
 $det \Phi \equiv \epsilon^{i_1,...i_{N_f}}
  \epsilon_{\bar{j}_1,...\bar{j}_{N_f}} \Phi_{i_1}^{\bar{j}_1}...\Phi_{i_{N_f}}^{\bar{j}_{N_f}}$.  Notice that it is therefore invariant under an anomaly-free discrete 
 $Z_{2N_f}$-subgroup of $U(1)_A$. 
  
  For spontaneous breaking of $U(1)_A$ and delivering a Goldstone boson 
  $\eta'$ with anomalous coupling  the necessary condition is that 
 the VEV  $\langle  det \Phi  \rangle$ is non-zero. \footnote{This can be seen by the 
 following simple argument. Imagine that $\Phi_{i}^{\bar{j}}$ has a VEV, but such that 
$\langle  det \Phi  \rangle = 0$.  Then, without loss of generality we can always make the VEV real by an anomaly-free  $SU(N_f)_L \otimes SU(N_f)_R$ transformation.
Thus, we can eliminate all the phases without $U(1)_A$-rotation and hence 
no $\eta'$ will emerge with anomalous coupling. }

  If this were to be the only available non-zero order parameter,
 $\langle  det \Phi  \rangle \neq 0$,  the flavor symmetry would be broken down to 
 $SU(N_f)_L \otimes SU(N_f)_R \otimes U(1)_V$. 
  
   We need to understand what happens to the rest of the flavor group.  
 Let us now show that the flavor symmetry  $SU(N_f)_L \otimes SU(N_f)_R$ must be spontaneously broken down to its anomaly-free subgroup.

\subsubsection{Absence of massless fermions  at low energies: Gravity as a spectator } 

 As a first step towards this proof, let us convince ourselves that 
 in the low energy theory of Bob there is no place for any massless fermions.

  We shall try to prove this by contradiction.  Let us assume that such massless fermions do exist and denote them by $\lambda_{\alpha}, ~ \alpha = 1,2,...n$, 
 where $n$ is unspecified.  
   For definiteness, assume all of them to be written in left-handed Weyl basis.  Since we are making no assumption about the confinement, $\lambda$-fermions can be either a sub-set of  Alice's $\psi$-fermions 
  or their massless composites.  We wish to convince ourselves  that this set   is empty. 
  
   As the first step, notice that if $\lambda$-s exist they must form an anomaly-free 
   set with respect $U(1)_A$.  In order to see this  we  shall weakly gauge
    the $U(1)_A$-symmetry by introducing a gauge field $X_{\mu}$, which under the local version of the axial transformation (\ref{axial}) shifts as 
 \begin{equation}    
  X_{\mu} \rightarrow X_{\mu}\,  + \, { 1 \over g_X} \partial_{\mu} \alpha(x)\,, 
  \label{gaugedX}
  \end{equation} 
 where $g_X$ is the gauge coupling.  
   The gauged version of $U(1)_A$-symmetry we shall denote by
 $U(1)_X$, in order to distinguish it from the global axial symmetry. 
 
  This gauging produces gauge chiral anomaly. In order to cancel it we shall introduce 
 a Green-Schwarz axion, $a$,   that under the gauge transformation 
 (\ref{gaugedX})
 shifts as
 \begin{equation}      
  {a \over f_a}   \rightarrow  {a \over f_a}  \,  - \, \alpha(x)  \,.   
 \label{AxionX}  
\end{equation} 
Here $f_a$ is  a decay constant, which we shall later take to infinity. \\
 
 The  axion $a$ should not be confused with Weinberg-Wilczek axion \cite{axion}, 
 which is usually invoked for solving the strong-CP problem via Peccei-Quinn
 mechanism \cite{PQ}.  Rather, the axion $a$ is the axion of Green-Schwarz type \cite{GS}, the role of which  is the cancellation of gauge $U(1)_X$ anomalies.  
   The CP-invariance in our case is maintained by $\eta'$, as it was discussed 
   in details above. In this sense, in our setup the role of 
   Weinberg-Wilczek axion \cite{axion} is taken up by $\eta'$.  
   We shall explicitly show below, how 
   $a$ and $\eta'$ share their jobs of canceling anomalies and maintaining 
   CP-invariance.  \\

  The resulting gauge invariant Lagrangian of Alice has the following form  
   \begin{eqnarray}
  % \begin{equation}
   \label{AliceLX}
L_{Alice} \, &=& \,  \sum_{\psi} i \overline{\psi} \gamma^{\mu}  D_{\mu} \psi \,  -\nonumber\\  
 && -  \, F_X^2 \,  - \,  F^2  +  \, \theta F\tilde{F}  \nonumber\\
  && + \, (f_a)^2\left (g_XX_{\mu}  + \partial_{\mu}{a \over f_a} \right )^2 \, +    
  \, \nonumber\\  
 && + \, {a \over f_a} \left (F_X\tilde{F}_X  \, + \, F\tilde{F} \right ) \,. 
 \end{eqnarray}
  The first row describes the gauge invariant coupling of $\psi$-fermions to 
 gluons and to $X_{\mu}$, with $D_{\mu}$ denoting a standard  
 covariant derivative with respect to both interactions. 
  The contraction of color and flavor indexes is obvious and is not written explicitly.  
  The second row describes the usual gauge field strengths, with irrelevant constants absorbed, plus the $\theta$-term for QCD. The $\theta$-term 
  for the abelian group is  irrelevant. 
   The third row describes the gauge invariant mass of $X_{\mu}$, in which $a$ plays the role of the St\"uckelberg field. 
  Finally, the fourth row describes the coupling of the axion to the  dual field strengths.  
 The normalization is such that under  $U(1)_X$  the fermions contribute  an anomalous shift, 
 \begin{equation} 
 \delta L_{fermion}  = \alpha \left (F_X\tilde{F}_X  \, +  F\tilde{F} \right )\,,  
 \label{fermionSX}
 \end{equation} 
 which is exactly compensated by the shift of axion (\ref{AxionX}).  \\
 
   Notice, if we switch for a moment to the conventional language of explicit symmetry-breaking by instantons, we will need to 
 add the 't Hooft determinant term to  Alice's Lagrangian.  
  This term has a form 
   \begin{equation} 
 L_{'t Hooft}  \, = \, e^{i \left ({a \over f_a} - \theta \right)} det(\bar{\psi}\psi)\, + \,... \, ,   
  \label{detFermion1} 
  \end{equation}
 and is manifestly gauge invariant under  $U(1)_X$. This shows clearly that 
 - as was stressed in the introduction - the instantons {\it cannot}  avoid requirement of anomaly matching for $U(1)_X$-symmetry. This constraint was not taken into the account in 't Hoofts original paper \cite{anomalymatching}.\\

 Notice, that on top of the anomaly-free local symmetry, there continues to exist the 
 ``old" anomalous global symmetry $U(1)_A$, which only acts on fermions and not 
 on either $X_{\mu}$ or $a$.  Because of this, the  existence of new gauge symmetry $U(1)_X$ is not affecting the solution of the strong-CP problem: 
 Alice can still shift away the $\theta$-term by the old anomalous global chiral rotation.  This is also clear from the fact that the new degrees of freedom that we have added to the theory can be made arbitrarily massive, by taking 
 the quantity $m_a \equiv g_Xf_a$ large.  Below the scale 
 $m_a$  the modes  $X_{\mu}$ and $a$ decouple and can be integrated out. 
 Alice's theory then is left with Anomalous global symmetry $U(1)_A$ as before.  
 The same continues to be true for arbitrarily small, but finite, value of $m_a$.  
 
 Correspondingly,  in the low energy theory of Bob the story with $\eta'$  
persists as before, modulo a small mixing with $a$.    By taking the couplings $g_X$ and $1/f_a$ small, we can 
make  $X_{\mu}$ and $a$ to interact with the QCD sector  arbitrarily weakly. In this way they only play the role of spectators for monitoring the anomaly matching. \\    

 Now, first we need to show that in  Bob's theory, in the absence of $\lambda$-fermions, 
 the gauge anomaly is fully matched by  $\eta'$, and that the decoupling limit,
 \begin{equation} 
 g_X  \rightarrow 0,~~f_a \rightarrow \infty, ~~m_a=finite, 
 \label{decouplingX} 
 \end{equation} 
 is smooth.    
  The relevant part of Bob's theory without $\lambda$-fermions
 has the following form, 
 \begin{eqnarray}
 \label{BobLX}
  % \begin{equation}
 L_{Bob} (bosons) \, &&= \, {1 \over 2}  E^2 \, - \,\left (  {\eta' \over f_{\eta}}  + 
  {a \over f_a} \right) E \, -  \nonumber\\   
&&  -  \,  F_X^2 \, + \, {1 \over 2} (f_a)^2\left (g_XX_{\mu}  + \partial_{\mu}{a \over f_a} \right )^2 \, +    
  \, \nonumber\\  
 &&  + \,  {1 \over 2}  (f_{\eta})^2\left (g_XX_{\mu}  - \partial_{\mu}{\eta' \over f_{\eta}} \right )^2     \,,  
 \end{eqnarray}
 (where we have restored some numerical factors for convenience). 
The relative signs inside the brackets are important, since they ensure the gauge invariance under the above-discussed normalization.  The gauge-invariance of the theory under  $U(1)_X$ is ensured by the corresponding shift of $\eta'$:
 \begin{equation}      
  {\eta' \over f_{\eta} }   \rightarrow  {\eta' \over f_{\eta} } \,  +  \, \alpha(x)  \,.   
 \label{ETAX-shift}  
\end{equation} 
This shift is fixed from the fact that $\eta'$ transforms as the phase of the fermion 
't Hooft determinant.   

  It is clear that although  spectators ensure the invariance of the theory under 
  the local $U(1)_X$-transformation, they do not affect  Bob's solution of the strong-CP 
  problem, and the decoupling limit (\ref{decouplingX})  is smooth.

Indeed, solving the equation of motion for $E$ we obtain, 
\begin{equation}\label{EX}
E \, = \, \left( {\eta' \over f_{\eta}}  + 
  {a \over f_a} -\theta \right),
\end{equation}
where $\theta$ as before is an integration constant that is matched with Alice's 
$\theta$.  
We see that the electric field $E$ continues to be nullified by the following 
degree of freedom, 
\begin{equation}\label{Etilde}
\tilde{\eta'} \, \equiv  \,  { f_a\eta'  + f_{\eta} a   \over \sqrt{ f_a^2 + f_{\eta}^2}}\,.
\end{equation}
 Diagonalizing the mass-matrix, we can see easily that it is exactly this degree of freedom that is eaten-up by the QCD $3$-form $C$ and is getting a mass, 
\begin{equation}
m_{\tilde{\eta'}}^2 \, =   \,  {1 \over  f_a^2 }  + {1 \over  f_{\eta}^2} \, .  
\label{massEX}
\end{equation}  
 The orthogonal combination, 
 \begin{equation}\label{Atilde}
\tilde{a} \, \equiv  \,  { f_{\eta'} \eta'  -  f_{a} a   \over \sqrt{ f_a^2 + f_{\eta}^2}}  ,
\end{equation}
 is eaten up by the $X_{\mu}$ gauge field and  is getting the following  mass, 
 \begin{equation}
m_{\tilde{a}}^2 \, =   \,  g_X^2 \, \left (f_a^2 + f_{\eta}^2 \right ) \,. 
\label{massEX}
\end{equation}  
That is, $\tilde{a}$ becomes a longitudinal polarization of  $X_{\mu}$ and the 
two form a Proca field of mass $m_{\tilde{a}}$. \\ 

[As a consistency check, notice that $\tilde{\eta'}$ is exactly the combination 
that would get mass from  't Hooft determinant (\ref{detFermion1}), if we were 
to use the conventional language of instantons.  This is obvious from plugging the 
last expression in (\ref{detFermion}) into (\ref{detFermion1}).  
This clearly shows that the two languages are alternatives  to each other and must not be used simultaneously, in order to avoid double-counting.] \\

  In the decoupling limit (\ref{decouplingX}) 
 % $g_X \rightarrow 0, f_a\rightarrow \infty$, 
  we have 
  \begin{equation}
  \tilde{\eta'}  \rightarrow \eta',~{\rm and}~ 
   \tilde {a} \rightarrow  a \,, 
 \label{limitX}
\end{equation}     
 and the two sectors decouple. Thus, the limit is smooth and the  strong-CP problem is solved for arbitrary values of $g_X$ and $f_a$.   The anomalous shifts generated by $\eta'$ and $a$ exactly cancel each other for arbitrary choice of these parameters.   
 
  From here it is clear that $\lambda$-fermions must be anomaly-free
with respect to  $U(1)_X$. Correspondingly, they must be anomaly-free 
under the global $U(1)_A$-symmetry as well, since the two act on fermions 
in the same way. \\
 
  Let us now show that  from the last statement it follows that $\lambda$-s cannot
  exist, because  them being  anomaly-free with respect to  $U(1)_A$, 
 contradicts to the anomaly matching 
  with respect to another spectator interaction that we shall now introduce - {\it gravity}. \\

   We thus, take into the account coupling to  gravity.  We shall treat gravity as a spectator interaction, keeping in mind a limit of  infinitesimally small Newtonian coupling.  
   
   Notice, since $\lambda$-s are massless, there always exists a chiral symmetry 
  \begin{equation}
   \lambda \rightarrow e^{i\beta} \lambda \,, 
   \label{YLambda} 
   \end{equation}
   which is anomalous with respect to gravity.  We shall gauge this symmetry 
   and call it $U(1)_Y$.  The important thing is that 
   $U(1)_Y$ must be  different from $U(1)_X$, since, as we just concluded, under the latter 
   $\lambda$-s are anomaly-free.

   The  gauge field of $U(1)_Y$ we shall denote by $Y_{\mu}$.  This chiral symmetry exhibits both $(U(1)_Y)^3$ as 
  well as a mixed $U(1)_Y$-gravity-gravity anomaly \cite{gravityanomaly},  due to which the Lagrangian shifts as, 
      \begin{equation}
      \label{RRgravY}
\delta L_{Bob} \, = \, \beta(x)  (R\tilde{R} + F_Y\tilde{F}_Y) \, .
 \end{equation} 
 Once again, we absorb the anomaly coefficients into the normalizations  of 
 $R\tilde{R}$ and $F_Y\tilde{F}_Y$.  In order to cancel the above anomalies we introduce a new  
 spectator Green-Schwarz axion $b$, with a decay constant $f_b$. 
The $U(1)_Y$-gauge transformation acts on $Y_{\mu}$ and $b$ as 
 \begin{equation}    
  Y_{\mu} \rightarrow Y_{\mu}\,  + \, { 1 \over g_Y} \partial_{\mu} \beta(x)\,,~~ 
   {b \over f_b}   \rightarrow  {b \over f_b}  \,  - \, \beta(x)  \,.   
 \label{AxionY}  
\end{equation} 
 where $g_Y$ is the gauge coupling. 
  
  The gauge invariant version of fermionic part of Bob's Lagrangian becomes, 
   \begin{eqnarray}
     \label{BobFermion}
&& L_{Bob}(fermi) \, = \,  \sum_{\lambda} i \overline{\lambda} \gamma^{\mu}  D_{\mu} \lambda \,  -  \, F_Y^2 \,  + \nonumber\\ 
  && + \, (f_b)^2 \left (g_YY_{\mu}  + \partial_{\mu}{b \over f_b} \right )^2 \, +    
   \, {b \over f_b} \left (F_Y\tilde{F}_Y  \, + \, R\tilde{R} \right )  \, +  \nonumber\\
 && + \, ({\rm  pure ~ gravitational~action})  \, ,  
 \end{eqnarray}
where, of course derivatives are replaced by their covariant versions  with respect to all gauge interactions including gravity.  Obviously,  the gravitational anomalies 
 of $U(1)_X$ are cancelled among $\eta'$ and $a$ and that sector is not displayed 
 explicitly.   

Notice, in case if  Bob's action contains some higher order interactions 
among $\lambda$-s that break the global  $U(1)_Y$-symmetry explicitly down to some 
discrete $Z_n$-subgroup, with $n$ some integer,  this is no obstacle for gauging it. 
  We just need to promote the coefficient of every such operator into a
  {\it spurion}  that transforms under gauge  $U(1)_Y$ in the appropriate  way. 
  For this the simple prescription is:  Each fermionic  field $\lambda$     
 or $\bar{\lambda}$ entering such a vertex, must be accompanied by 
 a factor $e^{i{b \over f_b}}$ or $e^{-i{b \over f_b}}$ respectively.  For example, an operator
 $\lambda^n e^{i{b \over f_b}n}$ is manifestly gauge invariant. 

 Now let us move into the theory of Alice.  All the spectator interactions including 
 gravity, $Y_{\mu}$ and the axion $b$, continue to be part of Alice's theory, but - by definition - not 
 the $\lambda$-fermions.  Under the gauge $U(1)_Y$-transformation  $b$
 continues to generate the anomalous shifts of the Lagrangian, which 
 would be matched by (\ref{RRgravY}) if $\lambda$-fermions were around.      
  
  Thus, we are ready to ask:  What matches the $U(1)_Y$-anomalies
generated by the axion $b$  in theory 
 of Alice?  Alice only has $\psi$-fermions at her disposal.  But, the only symmetry anomalous with respect to gravity that acts on $\psi$-s is axial $U(1)_X$. 
 This creates an inconsistency: First, 
  we know that $U(1)_X$  and $U(1)_Y$ must be different;  Secondly, 
all the $U(1)_X$-anomalies, including the mixed gravitational one,  are fully taken care by 
 the axion $a$.  So in Alice's theory there are no fermions that could compesate the gravitational anomaly of $U(1)_Y$. \\
 
 Thus, the assumption that in Bob-s theory there can exists 
 some massless $\lambda$-fermions lead us to a contradiction.   
Thus, there  is no room for any massless fermions (composite or elementary) in the theory of Bob. \\
    
 Notice, for reaching this conclusion is was important to use gravity as a spectator interaction.   Without gravity, we could only prove that the low energy massless 
 fermions must be color singlets, but would not be able to exclude their existence. 
 The power of gravity is that there are no ``gravity-singlets" in nature and 
 existence gravitational anomaly is universal for any set of  massless fermions.   
This gave us an additional matching condition, which is absent in pure QCD.

 \subsubsection{Breaking of  $SU(N_f)_L \otimes SU(N_f)_R$ Chiral Symmetry}

      As  a result of the above analysis,  we are left with the following two logical possibilities: \\
   
    {\it (1)}  The fermions stay elementary, but acquire effective mass-terms from symmetry breaking; \\
   
    or  \\
    
    {\it (2)} The massless fermions disappear  at low energies, because they all form bound-states. Since we have already excluded the possibility of massless fermionic 
    bound-states,  any bound-state formed by the fermions must be    
 either massive or bosonic.  
    The formation of the bound-states may or may not be due to confinement, about which 
    we make no assumption. 
     \\
    
     The above two logical possibilities are not inter-exclusive. 
     We need to prove that each of of them implies that the chiral symmetry is completely broken.  \\

  {\it  Option  (1)}.  From the  option {\it (1)} the complete breaking of chiral symmetry 
follows in an obvious way.  It is enough  to notice that the fermion mass matrix $M_{\bar{j}}^{i}\, \bar{\psi}^{\bar{j}} \psi_{i}$
 transforms as a bi-fundamental representation $(\bar{N}_f, N_f)$
 of  $U(N_f)_L\times U(N_f)_R$ flavor symmetry.  For an arbitrary  bi-fundamental  
 matrix $M_{\bar{j}}^{i}$ with all the eigenvalues non-zero the maximal little group is $SU(N_f)_{R+L} \otimes U(1)_V\otimes Z_{2}$.  Hence, this is the largest sub-group that can survive unbroken, under the condition that all fermions are massive.  \\
 
 {\it Option (2)}. The option {\it (2)} also implies breaking of 
$SU(N_f)_L \otimes SU(N_f)_R$, at least  down to its maximal anomaly-free subgroup.

 The proof, follows from applying 't Hooft's anomaly matching condition
 with spectator fermions \cite{anomalymatching}. 
  The reason in our case it allows us to make a stronger statement 
is because we have already gained a small advantage with respect to 't Hooft's starting position: 
 Thanks to the matching anomalies of weakly-gauged axial symmetry,  while using gravity and Green-Schwarz axions as spectators, we have already eliminated the possibility of any massless fermions altogether.  This simplifies our task significantly. 

  The argument can be presented as a particular case of a more general situation. \\
 
  Consider a theory based on a global chiral symmetry group $G_{chiral}$ and with the set of fermions $\psi$ transforming under it.  We shall allow 
 $\psi$-fermions  to also interact  via  some other gauge interaction $H$. 
 Now, assume that, due to unspecified dynamics of $H$, below certain scale  $\Lambda$ all the $\psi$-fermions form bound-states, so that there are no fermions transforming 
 under $G_{chiral}$ in the low energy theory. 
  
  Then,  we can prove that in deep infrared the chiral group $G_{chiral}$ 
 must be spontaneously broken at least down to its maximal anomaly-free subgroup. \\

Following 't Hooft,  let us assume that $G_{chiral}$ is gauged weakly. 
 (In fact,  would be enough to gauge different sets of chiral abelian subgroups of $G_{chiral}$.) We do this by introducing 
  the set of gauge fields $B_{\mu}^{(m)}$ with $m = 1,2, ...$. 
  
    This gauging creates the chiral anomalies  
  of $(G_{chiral})^3$-type. In order to cancel them, we introduce the set of massless colorless chiral fermions 
  $\chi$ only charged under $G_{chiral}$.  Now, above the scale $\Lambda$ the  
  $(G_{chiral})^3$ anomalies are cancelled among $\chi$-s and 
  $\psi$-s. But, below the scale $\Lambda$, by assumption  there are no massless fermions available from the $\psi$-sector.  
  
   So, if the gauge fields $B_{\mu}^{(m)}$ of $G_{chiral}$ survive massless at low energies, it would be impossible to match the anomalies created by $\chi$-s. Thus,  the only consistent possibility is that below the scale $\Lambda$ the group $G_{chiral}$ is Higgsed  down to its anomaly-free subgroup, 
   $G_{free} \subset G_{chiral}$, by some order parameters formed out of $\psi$-s.  Since the argument goes through for arbitrarily-weak gauge coupling $g_B$, it also should persists for the case when $G_{chiral}$ is global.  
  
   Thus, we conclude  that even if all the $\psi$-fermions at low energies are in either massive or bosonic bounds-states, nevertheless the anomalous part of the chiral flavor group
  $G_{chiral}$ must be fully broken.\\
   
  It is trivial to apply the above general argument to QCD with $N_f$ flavors.  
  We just have to take  $G_{chiral} = SU(N_f)_L\times SU(N_f)_R$ and assume that 
  the role of $H$ is played by the gauge color group.  We then conclude that 
  the flavor group must be spontaneously broken down to an anomaly-free subgroup
  $G_{free}$. \\
  
  Let us reiterate that this is a stronger conclusion than what one would reach by  applying 't Hooft's anomaly matching  without knowing that massless fermions are forbidden at low energies.   In such a case, the  surviving low-energy 
group would not need to be anomaly-free.  It would only need to deliver the same anomaly as the high-energy one, in order to match the anomalies of the spectator $\chi$-fermions. 
 This would give a milder constraint. \\

  The physical meaning of the spontaneous breaking of 
 chiral flavor symmetry by the condensate of $\psi$-fermions is very transparent.   
 This dynamical symmetry-breaking is necessary, in order for $\psi$-fermions
 to deliver the Nambu-Goldstone bosons, the pions $\pi^{(m)}$, which 
 effectively assume the role of  St\"uckelberg-Wess-Zumino-Green-Schwarz type 
 axions and cancel the anomalies of $\chi$-fermions in Bob's theory.  
 
For example, consider a pion $\pi^{(m)}$ resulting from the spontaneous breaking of 
a given anomalous  abelian subgroup  $U(1)_{chiral}^{(m)}$.  
Integrating out $\psi$-fermions, this pion acquires an anomalous 
 couplings, 
 \begin{equation} 
   \label{BobBb}
L_{Bob} \, = \, {\pi^{(m)} \over f_{\pi^{(m)}}} F_{(m)}\tilde{F}_{(m)} \, + \, ... \, ,
 \end{equation} 
 where $F_{(m)}$ is the field strength of $B_{\mu}^{(m)}$ and $\tilde{F}_{(m)}$ 
 its dual (again, the coefficient has been rescaled in normalization).  
 
  Under,  $U(1)_{chiral}^{(m)}$ the pion shifts as 
   \begin{equation}    
    {\pi^{(m)} \over f_{\pi^{(m)}}} \,  \rightarrow   {\pi^{(m)} \over f_{\pi^{(m)}}} \,  +  \, \beta^{(m)}(x) \,.  
  \label{shiftPionQCD}
   \end{equation}
   and cancels $U(1)_{chiral}^{(m)}$-anomaly generated  by  the $\chi$-fermions
  in Bob's theory.

   This completes the argument. 
 
 \section{Generalization to gravity}
 
 Although in our discussion we used the QCD-terminology, 
 we tried to keep it maximally general and independent from the  
 specifics of the QCD dynamics. 
 
  Correspondingly,  we can attempt to apply our arguments to any theory which has a non-zero topological  susceptibility in the absence of fermions 
and a chiral anomaly in their presence. Every step of our discussion can be repeated to conclude that in any such theory the fermions must condense 
and break Chiral symmetry.  Moreover, there must exist a
pseudo-Goldstone analog of $\eta'$ that gets a mass in this process,  by being eaten-up  by  a $3$-form.  Beyond this point the specifics of the theory must be 
taken into the account and discussion becomes more subtle.

We shall now apply our reasoning to gravity. 
  It has already been noticed \cite{gia3form} that gravity 
contains both of the required ingredients.

 First, it contains  $C$-field in form of  a  gravitational Chern-Simons $3$-form, 
\begin{equation}
C_g\equiv \Gamma \wedge d \Gamma  + {2 \over 3}  \Gamma\wedge \Gamma\wedge \Gamma,~~~ 
E_g \equiv R\tilde{R}= ^*\mathrm{d}C_g\,,  \label{eqEG}
\end{equation}
where $\Gamma$ is the connection, $R$ is the Riemann tensor
and $\tilde{R}$ its Hodge-dual. 

Secondly,  the  chiral current  exhibits the well-known chiral gravitational 
anomaly \cite{gravityanomaly},  
\begin{equation}
      \partial^{\mu} J_{\mu}^{(A)} \, =  \, R\tilde{R} \, , 
    \label{divRR} 
    \end{equation}
where, the current includes all the fermions of the
theory $\psi_{(j)}$, which can be written in the same handiness basis (e.g., all left-handed).   This introduces a difference that in gravity, with the same number of fermion degrees of freedom,   
the flavor group is larger than in QCD.  Ignoring other interactions, the flavor group of gravity is  $U(N)$, where $N$ counts number of all the
massless Weyl  fermions.

   At present we have no knowledge about the existence of non-zero topological susceptibility of the vacuum in pure gravity (i.e., gravity without massless fermions 
   and/or axions).   Therefore, we take this as our starting {\it assumption}.  We  assume that due to some unspecified physics the following correlator is non-zero 
in a fermion-free version of quantum gravity,   
    \begin{equation}\label{ETGrav}
\langle E_g, E_g\rangle_{p\to 0}\, = \, const \neq 0  \, . 
\end{equation} 
   
     Einstein gravity exhibits no mass gap (i.e., graviton is massless). 
  However, as we have discussed, the existence of the constant electric field 
  $E$ in the vacuum of pure glue, directly follows from the assumption of non-zero 
  vacuum topological susceptibility and is independent from  the assumptions of mass gap and/or confinement.   
  
The same argument goes through in gravity:   From the assumption of non-zero vacuum topological susceptibility (\ref{ETGrav}), it
directly follows that  $C_g$ is a massless $3$-form field, with 
its propagator having a pole at $p^2=0$.  
 Correspondingly,  the vacuum houses a  constant CP-odd electric field 
 \begin{equation}
 E_g = \theta_g \,. 
 \label{Egrav}
 \end{equation} 
 Just like 
in case of pure glue,  this electric field is the way Bob understands the physical CP violation, which in Alice's description is due to  
physical observability of the gravitational $\theta_g$-term,
\begin{equation}
 \label{RRgrav}
L_{Alice} \, = \, \theta_g R\tilde{R} \, + \,  ...\, .
 \end{equation} 
  When Alice changes the value of $\theta_g$, Bob changes 
the value of $E_g$. \footnote{Some implications of the gravitational analog of 
$\theta$-term,  in theories with lower dimensions or on backgrounds with broken 
Poincare invariance, where previously discussed in \cite{CSgrav}. Our focus here is very different.  Since we assume non-zero topological susceptibility, the term 
(\ref{RRgrav}) is physical  without the need of violation of Poincare symmetry.} \\

  The next step is to introduce some massless fermions.  In order to make comparison  with the case of QCD, let us introduce the same number of fermionic degrees of freedom. 
What in QCD would be $N_f$ flavors, by gravity are seen as $N = 2N_f$ massless Weyl fermions, all of which  can be written in left-handed basis, $\psi_{i}$, where $i =1,2,...N$. These fermions form a fundamental representation of $U(N)$ flavor group.
The $U(1)_{A}$ subgroup of this symmetry,  which acts on the fermions as, 
   \begin{equation}
    \psi_{i} \rightarrow {\rm e}^{i\alpha}  \psi_{i} \,,     \label{axialG}  
  \end{equation}
is anomalous with respect to gravity.  The corresponding current  $J_{\mu}^{(A)}$ exhibits an anomalous divergence  \cite{gravityanomaly}  given by  (\ref{divRR}). 
Therefore, under the chiral transformation (\ref{axialG})  the Lagrangian shifts 
as 
 \begin{equation}
     \delta L_{Alice} \, =  \, \alpha R\tilde{R} \,.  
    \label{varRR} 
    \end{equation}

Once again,  all additional coefficients are absorbed in normalization of 
$R\tilde{R}$.     
 Due to this, just as in QCD, the introduction of massless fermions 
makes the gravitational analog of $\theta$ unphysical and correspondingly makes the vacuum topological susceptibility zero.  Correspondingly,  CP is unbroken. 
  The restoration of CP-invariance is {\it universal} and Alice and Bob must agree on this fact.  \\ 
  
  In  Alice's language this is understood as a freedom  
 to arbitrarily shift $\theta_g$ by  $U(1)_A$ chiral transformation (\ref{axialG}).  \\
 
 In Bob's theory, this must be matched by the appearance of  a pseudo-scalar 
 $\eta_g'$ that  Higgses the gravitational Chern-Simons $3$-form $C_g$
 exactly in the same way as $\eta'$-meson Higgses the analogous 
 Chern-Simons $3$-form in QCD.   The effective Lagrangian describing this phenomenon is fixed by the gauge invariance and the anomalous coupling, and 
 therefore,  is very similar to  (\ref{effectiveQCDeta})
%%%%%%%% 
 \begin{equation}\label{effectiveG}
L_{Bob} \, = \, {1 \over 2}  E_g^2 \, - \, {\eta_g' \over f_{g}}  \, E_g \, + \,  {1 \over 2} \partial_{\mu}{\eta_g'} \partial^{\mu} {\eta_g'},
\end{equation}
Of course, here and everywhere the generally-covariant contraction of indexes is assumed.  
Solving the equation of motion for $E_g$, 
\begin{equation}\label{solEG}
E_g \, = \, \left({\eta_g' \over f_{g}}-\theta_g\right),
\end{equation}
and plugging it into 
the equation for $\eta'_g$, we get, 
\begin{equation}\label{boxetaG}
\Box\, \eta'_g +  {1 \over f_{g} }\left({\eta'_g  \over f_{g}} -\theta_g \right) = 0 \,.  
\end{equation}
Thus, exactly as it happened in QCD, we conclude that the vacuum is at   ${\eta_g' \over f_{g}}=\theta_g$ so that the VEV of the CP-odd electric field vanishes, $E_g=0$.   This is the way Bob understands 
CP-conservation in his theory. 
   CP-invariance of the vacuum implies that under the chiral transformation 
   (\ref{axialG}) performed by  Alice, $\eta_g'$ must shift as   
        \begin{equation}
   {\eta'_g \over f_{g}} \rightarrow  {\eta'_g \over f_{g}} \, + \, \alpha \,, 
 \label{shiftEtaG}
\end{equation}
in order to compensate the shift of $\theta_g$,
        \begin{equation}
   \theta_g \rightarrow  \theta_g \, + \, \alpha \,,  
 \label{shiftThetaG}
\end{equation}
which follows from (\ref{varRR}).

 Thus, making the exact same matching - as was done in the case of QCD - between the descriptions of Alice and Bob,  
  we arrive to the conclusion 
 that  $\eta'_g$ is a pseudo-Goldstone boson, which originates from a spontaneous breaking of  
 $U(1)_A$ symmetry  (\ref{axialG}) triggered by a fermion condensate. 
 
   Introducing a notation $\Phi_{ij} \equiv \psi_i \psi_j$, we construct the 
 gravitational version of the order parameter for  breaking the $U(1)_A$-symmetry: 
 $det \Phi \equiv \epsilon^{i_1,...i_N}
  \epsilon^{j_1,...j_N} \Phi_{i_1j_1}...\Phi_{i_Nj_N}$, which is invariant under 
 the $SU(N)$ subgroup of the original  $U(N)$ flavor group.   
  
  Notice, $\Phi_{ij}$ being a Lorentz-invariant of Majorana-type 
 is symmetric in $i,j$-indexes and transforms as $N(N+1)/2$-dimensional representation 
  of $SU(N)$ flavor group. 
  For spontaneous breaking of $U(1)_A$ the necessary condition is that 
 the vacuum expectation value $\langle  det \Phi  \rangle$ is non-zero.   
 Existence of $\eta_g'$ indicates that this is the case. \\

  We now wish to understand what is the fate of $SU(N)$-part of the flavor symmetry. 
   
 For this we first need to check whether a conclusion that we reached in QCD 
 - namely, the absence of massless fermions in Bob's theory - holds in gravity. 
 In QCD we reached this conclusion by gauging the axial symmetry and matching anomalies. 
 We shall do exactly the same in gravity.  We promote global  $U(1)_A$ into 
 a gauge $U(1)_X$ symmetry and cancel anomaly by a Green-Schwarz axion 
 $a$.  Then, Alice's anomaly-free Lagrangian is,  
   \begin{eqnarray}
  % \begin{equation}
   \label{AliceXG}
L_{Alice} \, &=& \,  \sum_{\psi} i \overline{\psi} \gamma^{\mu}  D_{\mu} \psi \,  -\nonumber\\ 
 && -  \, F_X^2 \,   +  \, \theta_g R\tilde{R}  \nonumber\\
  && + \, (f_a)^2\left (g_XX_{\mu}  + \partial_{\mu}{a \over f_a} \right )^2 \, +    
  \, \nonumber\\  
 && + \, {a \over f_a} \left (F_X\tilde{F}_X  \, + \, R\tilde{R} \right ) \, + \nonumber\\  
 && +  \, M_P^2 {\mathcal R}  \,.    
 \end{eqnarray}
where the last row stands for pure-gravitational part with $M_P$ the Planck mass and ${\mathcal R}$ the Ricci scalar.  
Cancellation of gravitational anomalies works exactly as in the case of QCD: The would-be anomalous shift of the Lagrangian (\ref{varRR}) is 
compensated by the respective gauge shift of the axion (\ref{AxionX}).\\

 {\bf [} At this point we have to open a parenthesis, because  here comes the followings {\bf caveat}:  In QCD, both interactions - 
$U(1)_X$ and gravity - were introduced as spectators and both 
could be chosen to be {\it arbitrarily weak}.  
 In other words, there was no problem 
in taking the decoupling limit  (\ref{decouplingX}) smoothly, since simultaneously
we could take $M_P \rightarrow \infty$.  

But, in gravity the story is different, since gravity can not play the role of its own spectator. 
Correspondingly,  if we want to make sure that the decoupling limit is smooth 
we should be able to take  (\ref{decouplingX}) while keeping $M_P$ {\it finite}. 

 One may argue that this is not legitimate. There are the following two (inter-related)  reasons for that. 
 
 The first one follows from a wide-spread view of incompatibility between gravity and global symmetries.  Subscribing to this view, we have to be worried about taking
 the decoupling limit (\ref{decouplingX}) with finite $M_P$, since in this limit 
 we are left with a global $U(1)_A$-symmetry coupled to gravity.  
 
  The second reason is based on a stronger constraint that goes under the name of ``Weak Gravity Conjecture" \cite{weakgravity}. According to this conjecture,
  not only global, but also gauged symmetries are incompatible with gravity, if their strength is weaker than the strength of gravity.   
   
   Again, one could say that this argument precludes  us from taking  the  limit
(\ref{decouplingX}), while keeping a finite $M_P$. 
     
  However, given what we are after, 
%  not allowing ourselves to take the  decoupling limit,  
  the above would result  into a  circular argument 
 that would get us to nowhere.  We are not satisfied with a limited  knowledge that gravity is inconsistent with global symmetries and we want to understand why and how   
she gets rid of them.
 
 In other words,  our goal is precisely to understands how gravity responds to global symmetries dynamically.  For this, we must allow ourselves to take the global limit 
 and see what happens.  As we shall see, just as in case of QCD, gravity responds by eliminating massless fermions and breaking all the chiral symmetries dynamically. 
 
 In this respect, the results of the present paper can be viewed as a dynamical  
 justification  for incompatibility between global (or weakly gauged)
 symmetries with gravity:  If we try to insist on such global (or weakly-gauged) symmetry, the fermions condense 
 and break it {\it dynamically}! {\bf ]}\\
 
 With above in mind we allow ourselves to take $g_X$ and $f_a^{-1}$ arbitrarily small, while keeping  finite $M_P$.  
 
 The rest of the arguments closely follow the QCD reasoning. First, in theory of 
 Bob the axion shift is matched by $\eta_g'$.  The corresponding gauge invariant
 Lagrangian is just a gravitational version of (\ref{BobLX}),  
 \begin{eqnarray}
 \label{BobLXG}
  % \begin{equation}
 L_{Bob} (bosons) \, &&= \, {1 \over 2}  E_g^2 \, - \,\left (  {\eta_g' \over f_g}  + 
  {a \over f_a} \right) E_g \, -  \nonumber\\   
&&  -  \,  F_X^2 \, + \, {1 \over 2} (f_a)^2\left (g_XX_{\mu}  + \partial_{\mu}{a \over f_a} \right )^2 \, +    
  \, \nonumber\\  
 &&  + \,  {1 \over 2}  (f_g)^2\left (g_XX_{\mu}  - \partial_{\mu}{\eta_g' \over f_g} \right )^2  + \nonumber\\
 && +  \, M_P^2 {\mathcal R}  \,,    
 \end{eqnarray}
where the gauge shifts of $\eta'_g$ and $a$ compensate each other. 
  From here, performing exactly the same steps as in QCD, we can conclude that 
in Bob's theory there cannot exist any massless fermions, since chiral anomaly 
is fully booked by $\eta'_g$. \\

Indeed, if either a subset of Alice's $\psi$-fermions or their composites could survive 
massless in Bob's theory, this surviving massless set $\lambda$ must be anomaly-free with respect to $U(1)_X$, since this anomaly cancels among 
the axion $a$ and $\eta'_g$-meson. But, since $\lambda$-s are massless, there inevitably exists a chiral symmetry $U(1)_Y$  acting on
them as (\ref{YLambda}).  Just as in the case of QCD, we gauge this symmetry 
introducing the spectator gauge field $Y_{\mu}$ and a Green-Schwarz axion 
$b$, which transform under  $U(1)_Y$ as  (\ref{AxionY}). 
The axion $b$ cancels  $U(1)_Y$ anomalies of  $\lambda$-fermions. This sector of 
 Bob's Lagrangian has the form (\ref{BobFermion}).  So far so good.  But  
 now, going to Alice's theory, there are no fermions there able to cancel the 
  anomalous shift of $b$-axion.  Obviously, $\psi$-s cannot do this job, since 
  the only symmetry anomalous with respect to gravity under which they transform 
  is $U(1)_X$, which - by construction - is different from $U(1)_Y$.   
 
  Thus, we see that Bob's theory is not hospitable to any massless fermion species.

 This leaves us with the same two options as in QCD:  The fermions  
must either get masses from the chiral symmetry breaking or form 
some massive composites (or both). 
 
  Again, gauging various chiral subgroups of the $SU(N)$ group 
  and performing 't Hooft's anomaly matching, we can conclude that 
  chiral symmetry must be broken down to its anomaly-free subgroup. 
  The only  technical novelty  in the matching argument is that, 
  unlike QCD, we cannot introduce the {\it massless}  spectator $\chi$-fermions, since  all fermions interact gravitationally and this will simply be equivalent to the increase of  $N$. 
  Thus, we have to use  Wess-Zumino-Green-Schwarz-type axions for such anomaly matching. \\
 
  The spectator axions can be introduced in a straightforward way for the entire 
  $SU(N)$-symmetry or for its subgroups.  As an useful guide-line, one can think 
  of these axions as originating from integrating-out the set of left-handed
  $\chi$-fermions transforming as anti-fundamental representation of 
  $SU(N)$-group. Then, the $\psi_i$- and $\chi^i$-fermions together form an anomaly-free 
  set of the $SU(N)$-group.  Now assume that $\chi$-fermions get large masses 
  from the Yukawa couplings to the VEVs of some elementary scalars  that Higgs the entire $SU(N)$-group.  The Higgs scalars do not have any Yukawa couplings to the $\psi$-fermions, but only to $\chi_i$-s and some additional   
$SU(N)$-singlet fermions.   As a result $\chi$-fermions mix with the additional 
gauge singlet fermions and form heavy Dirac particles.   Whereas,  the $\psi$-fermions remain massless. 

The Higgsing of  $SU(N)$-group results into $N^2-1$  Goldstone bosons 
$b^{(m)}~~(m=1,2,...N^2-1)$, which are eaten up by the corresponding $SU(N)$ gauge bosons, 
$B_{\mu} ^{(m)}$, and become massive.

By a suitable choice of the VEV of the Higgs fields and the value of $SU(N)$ gauge coupling, 
we can make the $\chi$-fermions {\it arbitrarily heavy} and the gauge bosons 
{\it arbitrarily light}.   After integrating out the heavy $\chi$-fermions, the anomalies 
of $\psi$ fermions  are cancelled by the Goldstone bosons  $b^{(m)}$  via effective 
 Wess-Zumino-Witten terms \cite{WZW}. 
 
 In other words, the Goldstone bosons $b^{(m)}$ are the low energy representatives of the heavy $\chi$-fermions  ``delegated" for the anomaly matching.  We must stress, however, that since in our case we are finally taking the decoupling limit, we can introduce the compensating Goldstone-bosons $b^{(m)}$ directly, without any explicit reference to the heavy fermions.

Now, we know that below some scale $\Lambda$ the 
$\psi$-fermions must decouple, since the low energy theory of Bob cannot contain any massless 
fermions.  Because the scale $\Lambda$ is totally independent of $SU(N)$-gauge dynamics, we can make the masses of $SU(N)$-gauge bosons  arbitrarily smaller than $\Lambda$.   By consistency, the anomalies generated by the spectator 
Goldstone bosons $b^{(m)}$ must continue to cancel in effective theory below 
the scale  $\Lambda$. 
This is only possible if $\psi$-fermions delegate their own representative Nambu-Goldstone bosons, $\pi^{(m)}$, 
 for canceling the anomalies of spectator Wess-Zumino-Green-Schwarz  axions, 
 $b^{(m)}$.   
 Thus, the $SU(N)$-symmetry must be spontaneously broken by $\psi$-condensate
 down to an anomaly-free subgroup, resulting into a set of Nambu-Goldstone bosons, $\pi^{(m)}$. 
 We shall refer to them as gravi-pions. \\

  For completeness of presentation, and in order to see how $\pi^{(m)}$-s and $b^{(m)}$-s share their jobs, let us explicitly go through the story for some chiral abelian sub-group of $SU(N)$-symmetry. We denote it  by $U(1)_{(m)}$  and the corresponding generator by $Q_{(m)}$. This symmetry  acts on 
fermion species as 
\begin{equation} 
\psi \rightarrow e^{i Q_{(m)} \beta^{(m)}} \psi \, ,
\label{Qgravity} 
\end{equation}  
where $\beta^{(m)}$ is a transformation parameter. 
 Let us denote the corresponding gauge field by $B_{\mu}^{(m)}$.
 Since, the generator $Q_{(m)}$ has a zero trace,
  the mixed anomalies of the type $U(1)_{(m)}-U(1)_X-U(1)_X$ and 
 $U(1)_{(m)}$-gravity-gravity are absent.  
 The anomaly $U(1)_{(m)}-U(1)_{(m)}-U(1)_X$ will be taken care by the coupling of axion $a$.  So we have to take care of 
 $(U(1)_{(m)})^3$ anomaly. 
 For this we need to introduce the corresponding Green-Schwarz axion, $b^{(m)}$.  We normalize its shifts under  (\ref{Qgravity}) as
  % \begin{equation}
    \begin{eqnarray}    
  B^{(m)}_{\mu}  &&\rightarrow \, B^{(m)}_{\mu}  + \, { 1 \over g_m} \partial_{\mu} \beta^{(m)} (x)\,,  \nonumber\\
   {b^{(m)} \over f_m} \,  &&  \rightarrow  {b^{(m)} \over f_m}  \,  - \, 
  \beta^{(m)}(x) \, , 
  \label{gaugeBb}
   \end{eqnarray}
  %\end{equation} 
 where $g_m$ is the gauge coupling and $f_m$ is the axion decay constant. 
  The resulting gauge invariant Lagrangian of Alice has the following form  
   \begin{eqnarray}
  % \begin{equation}
   \label{AliceBb}
L_{Alice} \, &=& \,  \sum_{\psi} i \overline{\psi} \gamma^{\mu}  D_{\mu} \psi \,  - F_{(m)}^2  +  \,  \nonumber\\  
 && + \, (f_m)^2\left (g_mB_{\mu}^{(m)}  + \partial_{\mu}{b^{(m)} \over f_m} \right )^2 \, +  \, \nonumber\\   
 && + \,  \left( c_m {a \over f_a}\,  + \, {b^{(m)} \over f_m}  \right) F_{(m)}\tilde{F}_{(m)} \, +...\, .\,
 \end{eqnarray}
  where $D_{\mu}$ is covariant also with respect to 
 (\ref{gaugeBb}).  The rest of the terms denoted by ``..." are the same as in (\ref{AliceXG}).    The constant $c_m$ is a relative anomaly coefficient that remains 
 after we absorb all other constants in normalizations of the dual field strengths. 
 
  Then, the anomaly generated by the fermion transformation (\ref{Qgravity}) is cancelled by the respective shift of  axion $b^{(m)}$ given by
  (\ref{gaugeBb}). \\
  
   We now go to theory of Bob, where, as we  know, there are no 
   massless fermions.  However, the spectator Green-Schwarz axions 
   continue to generate anomalies by their gauge shifts  (\ref{gaugeBb}). 
    What degrees of freedom cancel these anomalies? 
   
   The answer is that the $\psi$-fermions that cannot penetrate into Bob's theory  in form of fermions, must condense and deliver 
  the  Nambu-Goldstone degrees  of freedom, which have anomalous couplings and 
 shift appropriately  under the gauge symmetry.    
   That is, the only possibility to cancel the anomalous shift
   generated by  (\ref{gaugeBb}) is by the shift of a Goldstone boson which comes 
   from the phase of a fermion condensate that breaks $U(1)_{(m)}$ spontaneously.
   This  Nambu-Goldstone boson plays the role analogous to pion in QCD. 
   
   The  relevant part of  Bob's Lagrangian is, 
     \begin{eqnarray}
  % \begin{equation}
   \label{BobBb}
L_{Bob} (\pi_{(m)}) \, &=& - F_{(m)}^2  + 
 \,  \left({\pi^{(m)} \over f_{\pi^{(m)}}} +  {b^{(m)} \over f_m}  \right) F_{(m)}\tilde{F}_{(m)} \, +  \nonumber\\  
 && + \, (f_m)^2\left (g_mB_{\mu}^{(m)}  + \partial_{\mu}{b^{(m)} \over f_m} \right )^2
  \,  + \nonumber\\  
  && +  \,   (f_{\pi^{(m)}} )^2\left (g_mB_{\mu}^{(m)}  - \partial_{\mu}{\pi^{(m)} \over f_{\pi^{(m)}}} \right )^2 \,  +   \nonumber\\  
&&  +\,  \ c_m {a \over f_a} F_{(m)}\tilde{F}_{(m)} \, +
  ...\, . 
 \end{eqnarray}
where  `` ..."  includes (\ref{BobLXG}) and $f_{\pi^{(m)}}$ is the gravi-pion 
  decay constant.  The gauge shift (\ref{gaugeBb}) is  compensated 
  by the corresponding shift of pion, 
   \begin{equation}    
    {\pi^{(m)} \over f_{\pi^{(m)}}} \,  \rightarrow   {\pi^{(m)} \over f_{\pi^{(m)}}} \,  +  \, \beta^{(m)}(x) \,.  
  \label{shiftPion}
   \end{equation}
   
   %%%%%%%
    We see that the degree of freedom, 
    \begin{equation}\label{Btilde}
\tilde{b}^{(m)} \, \equiv  \,  { f_{\pi^{(m)}} \pi^{(m)}   -  f_{m}  b^{(m)}  
 \over \sqrt{ f_m^2 +  f_{\pi^{(m)}}^2}}  ,
\end{equation}
 is eaten up by the $B_{\mu}^{(m)}$ gauge field and is getting the following mass, 
 \begin{equation}
m_{\tilde{b}^{(m)}}^2 \, =   \,  g_m^2 \, \left ( f_m^2 +  f_{\pi^{(m)}}^2 \right ) \,. 
\label{massBtilde}
\end{equation}  
 The orthogonal combination,      
   \begin{equation}\label{Etilde}
\tilde{\pi}^{(m)}  \, \equiv  \, 
{ f_{\pi^{(m)}}  b^{(m)}    +  f_{m}   \pi^{(m)} 
 \over \sqrt{ f_m^2 +  f_{\pi^{(m)}}^2}} \,,
\end{equation}
remains massless. 

   In the decoupling limit  $g_m \rightarrow 0,~f_{m}  \rightarrow \infty$   
  we have 
  \begin{equation}
  \tilde{\pi}^{(m)}   \rightarrow \pi^{(m)} ,~{\rm and}~ 
   \tilde{b}^{(m)}   \rightarrow  b^{(m)}  \,, 
 \label{limitB}
\end{equation}     
 and the two sectors decouple. 
   
    Thus, we see that  the anomaly matching by the spectator Wess-Zumino-Green-Schwarz axions shows that  $SU(N)$ flavor group must be 
 spontaneously broken by the fermion condensate down to its anomaly-free subgroup.    Since in principle $SU(N)$ allows for different inequivalent embeddings 
 of anomaly-free subgroups, we cannot say with certainty down to which one the 
 symmetry must be broken.   The $SO(N)$ subgroup could be one natural outcome
 resulting in total $N^2/2 + N/2 -1$ Goldstones. \\
 
  For phenomenological considerations the $SU(48)$ flavor group would be of particular 
  interest, as this is the flavor group of  Standard Model fermions in the 
  limit in which all interactions except gravity are switched off.   This  flavor group naturally 
 admits the gauging of a  ``diagonal" anomaly-free $SO(10)_{G}$ subgroup defined by the following
 sequence of embeddings, 
  \begin{eqnarray}
  SO(10)_{G} &&\subset  SO(10)\times SO(10)\times SO(10) \subset \nonumber\\
  \subset  && SU(16)\times SU(16)\times SU(16) \subset SU(48) \, . 
  \label{SO10}
  \end{eqnarray} 
  The $48$-plet of fermions then reduces to three copies of $16$-dimensional spinor representation of $SO(10)_{G}$, which can serve as a symmetry group 
  of grand unification.

 \subsubsection{Persistence of dynamical symmetry breaking in presence of high-dimensional operators} 
   
  So far we were ignoring other sources of symmetry breaking. In gravity we cannot exclude  existence of high-dimensional operators that could break 
   $U(N)$-flavor symmetry explicitly.  Can a presence of such operators change our conclusion about the dynamical symmetry breaking? 
   
 The answer is negative, since such operators do not disturb anomaly matching 
 by axions. This  is because they are no obstacle for  gauging the relevant flavor symmetries. Using the corresponding Green-Schwarz axions,  such operators can be completed into their gauge-invariant versions.  After performing the anomaly matching, and establishing the fact of dynamical spontaneous symmetry breaking, 
we can decouple axions and recover back the operator that adds an  explicit breaking of a {\it spontaneously-broken}  global symmetry. This contribution generates non-zero masses to the corresponding pions, promoting them into 
pseudo-Goldstone bosons, but is not effecting their existence. 
 It is enough to demonstrate this for the chiral abelian subgroups. 
 
    The prescription is the following. Consider an arbitrary high dimensional operator 
  ${\mathcal O}( \bar{\psi},\psi)$ that  depends on fermions and their conjugates
  and breaks the flavor symmetry explicitly.  Let the charges of this operator under 
the set of chiral abelian symmetries $U(1)_{(m)} \times U(1)_X$ be 
$(q_m, q_X)$, in the units of charges of corresponding Green-Schwarz axions. 
 Then, this  operator can be easily promoted into its gauge invariant version  under these symmetries by means of multiplication by an appropriate exponential factor, 
 \begin{equation} 
 e^{- i \left (\sum_m q_m {b^{(m)} \over f_m }  + q_X{a \over f_a} \right )}
 {\mathcal O}( \bar{\psi},\psi) \, .
 \label{Ogauge} 
 \end{equation}  
   With this arrangement, we make sure that the gauge anomaly matching
  condition holds for arbitrarily-small values of the couplings between the spectator  
 axions and fermions.  Thus,  the previous conclusion  - that the anomaly-matching demands the existence  of  pions  in theory of Bob - is unaffected by high-dimensional operators.
In the decoupling limit, $f_m,f_a \rightarrow \infty$, we are left with global flavor symmetries that are, on one hand, spontaneously broken by the fermion condensate and, on the other hand, are explicitly broken by the high-dimensional operators (\ref{Ogauge}).  
   
      Correspondingly, the phenomenon of dynamical symmetry breaking by fermion condensate persists also in the presence of high-dimensional operators.  
      The only important effect of such operators in that they contribute to the masses 
      of  those gvari-pions on which the depend explicitly.  
      
   In other words,  masses of gravi-pions  become non-zero if 
    there exist operators that break corresponding generators of the global  flavor group  explicitly.  This masses however will be suppressed by the ratio of scales 
    of the fermion condensate and the scale of the high-dimensional operator, which 
  is unknown, but can be as large as the Planck mass. \\

 In summary, we showed that  under the assumption of non-zero topological susceptibility in fermion-free version of gravity,  after introduction of fermions with zero bare masses, the following happens:

\begin{itemize}

\item Fermions condense and form a non-zero order parameter 
that breaks the chiral  $SU(N)\times U(1)_A$-symmetry spontaneously
down to its anomaly-free subgroup.

\item The would-be Goldstone boson of 
$U(1)_A$, which we call $\eta_g'$,  becomes massive by 
being  eaten by a three-form $C_g$, via the effect closely analogous to Higgs phenomenon \cite{gia3form}.  This ensures the CP-invariance of gravitational 
vacuum.

 \item  No massless fermions exist in the low energy theory of gravity.  
 
  \item  The spontaneous symmetry breaking $SU(N)$ down to its anomaly-free 
  subgroup results into Goldstones, which we called gravi-pions. 
  The breaking to the largest anomaly-free subgroup $SO(N)$
 results into $N^2/2 +N/2 -1$ Goldstones.  
  
  \item  There may exist  some  high-dimensional operators that explicitly break 
  $SU(N)$ flavor symmetry. These can generate masses for gravi-pions. 
   However, they abolish neither  the phenomenon of dynamical symmetry breaking by fermion condensate, nor the generation of the fermion mass gap.

\end{itemize}

    Possibility of dynamical symmetry breaking by gravity is striking.  
  In the next section we shall try to understand the fundamental meaning of this phenomenon.

  \section{The role of micro black holes?}

    We have seen that the three phenomena -  breaking of chiral symmetry, generation of the  $\eta'$-meson mass and decoupling of fermions from the low energy theory -  follow from a single assumption of non-zero  topological susceptibility of the vacuum in fermion-free version of the theory, without any explicit assumption about the confinement.

    Applying this reasoning to gravity, we are lead to a rather striking conclusion that gravity responds to the existence of massless 
     fermions by dynamical breaking of chiral symmetry and generation of  mass gap
     in fermion spectrum.  \\

  This raises the following question: \\ 
  
  What underlying physics gives a non-zero 
  correlator  (\ref{ETGrav})  in  pure gravity, and why is it incompatible with the existence of massless fermion and chiral symmetry? \\
  
We think that the answer may be provided by  black holes
and this connection can shed light on another known puzzle in gravity.  

It is well-known 
that black holes are incompatible with continuous global symmetries. 
This knowledge often goes under the name of ``folk theorem",  which says that continuous global symmetries must either be gauged or broken.  The argument, 
that will not be repeated here,  relies on  no-hair properties of classical black 
holes.

But the question is: How the black holes manage to explicitly break the symmetry at the microscopic level?

  If we look closely at our previous analysis,  we may have a partial answer to this question. 
  At least the part that concerns the chiral symmetry. 
 Indeed, existence of massless fermions  coupled to gravity, would 
 imply existence of $U(N)$-flavor symmetry.   This would be incompatible with black holes.   But, we have shown that microscopic gravity could take care of this 
 potential inconsistency via topological susceptibility:  It generates
 a fermionic mass gap and simultaneously breaks the chiral symmetry dynamically.    

  The ``bootstrap" can be provided by the black holes themselves. Namely,   
 by  their contribution into the topological susceptibility of the vacuum. 
 
   The small black holes are natural candidates for contributing 
  into the correlator \cite{DFneu}.       
  Once we accept that black holes states are inevitable part of the gravity spectrum, then, everything shall fall into places, provided we assume that the micro-black hole states $|BH_k\rangle$ contribute into the vacuum correlator 
     \begin{equation}\label{BH}
\langle R\tilde{R}, R\tilde{R}  \rangle_{p\to 0}\, = 
\sum_k  \langle R\tilde{R}  |BH_k\rangle \langle   BH_k| R\tilde{R} \rangle  \, + \, ... \neq 0  \, . 
\end{equation} 
  This contribution provides the link between the black holes 
  and the breaking of global chiral symmetry.  \\

   Of course, without more detailed control of quantum gravity, we cannot compute the value of the correlator (\ref{BH}), however,  some reasonable guesses 
can be made.  
   
 First, it is reasonable to expect that the dominant contribution in (\ref{BH}) 
comes from {\it quantum}  black holes, i.e., the microscopic  black holes of the smallest size and mass.  Their existence follows from the existence 
of macroscopic  (classical) black holes. The simple way to see this is to wait until the black hole becomes so small that  its evaporation rate  becomes highly  non-thermal.   Deviations from thermality are set by inverse entropy \cite{giaThermal}, 
and therefore we can expect that the contribution into (\ref{BH}) 
from the black holes with smallest  entropy costs less suppression.  
Simultaneously, the corresponding size of a smallest entropy black hole marks the scale below which the quantum-gravity becomes strong.  

    Next,  we can give arguments why the scale of the correlator can naturally be smaller than the Planck scale, $M_P$.       
   In theory with $N$ particle species  
 there is an upper bound on the size of semi-classical  black holes, which is  given by the length $ = \sqrt{N} M_P^{-1}$ \cite{giaN}.   So, in such theories the quantum gravity scale is set by $M_P/\sqrt{N}$ rather than by $M_P$. 
 Secondly,  we should take into account 
the non-perturbative suppression, which could  be as strong as $e^{-N}$.

  What is interesting, is the universal nature of fermion condensate.  
  Since the gravitational chiral anomaly involves all the existing fermion species, the generation of fermion condensate and chiral symmetry breaking must be universal
  throughout all the fermions of the theory. 
  
  Of course, the most sensitive to the effect are the light fermions, and in the case of the Standard Model, the neutrinos.   So it makes sense to ask whether such a condensate can be a significant contributor into neutrino masses \cite{DFneu, Lena}.

   On the phenomenological side, the universal nature of the effect suggests  that the experimental lower bound on the masses of the lightest fermions  give us an experimental knowledge of the topological structure of the gravitational vacuum.  In particular, they give upper bound on the value of  gravitational vacuum topological  susceptibility somewhere not much above the neutrino mass scale.  
  
  \section{Relation with confinement?}
  
  The natural question to ask is whether there is any relation between our
  observations and phenomenon of the confinement, about which we have made no assumption. \footnote{In fact, if this assumption is added, the proof of chiral symmetry breaking in QCD becomes simpler.}  
Of course,  this assumption cannot apply to gravity. 
  
   Naively, it seems that our arguments diminish the role of confinement in the story of chiral symmetry breaking, as they show that solely the assumption of vacuum topological susceptibility suffices to do the job.  
  However, this is a superficial view.  We never proved that topological susceptibility of the vacuum can be non-zero in a non-confining theory. 
The only example of $3+1$-dimensional Poincare invariant  theory for which we are more or less confident that this quantity is non-zero, is massless-quark-free version of QCD, which is believed to be confining.  So, it could very well be 
that there is an intrinsic connection between this vacuum correlator and confinement.   
 Of course, having confinement as a necessary condition for its existence, would eliminate the possibility of non-zero gravitational topological susceptibility of the vacuum.  Obviously, in such a case our  arguments for  chiral symmetry breaking   would not apply to gravity. 

  What speaks against such a possibility is a seemingly-independent 
  ``folk theorem" argument, showing that black holes cannot tolerate chiral symmetry. 
 As discussed above, this cries for a ``bootstrap" scenario in which the black holes 
 make this intolerance self-consistent via breaking the chiral symmetry precisely by contributing into the gravitational topological susceptibility of the vacuum.  

  In this light, it would be extremely important to better understand the contribution 
  from black hole states in (\ref{BH}). The positive result would create a 
  precedent of breaking chiral symmetry in non-confining theory.  This would 
  also give reason for rethinking the role of confinement in chiral symmetry breaking also in confining theories, such as QCD.

\section*{Acknowledgements}

 We thank Cesar Gomez, for many lively and stimulating discussions,   
Gerard 't Hooft,  for discussions on anomaly matching using spectator axions and gravity, and Nico Wintergerst for useful discussions.   
  We thank Gabriele Veneziano, for pointing out ref. \cite{veneziano}
 and for discussions.  
   We thank Lena Funcke for discussions and collaboration on related phenomenological topics. 
This work was supported in part by the Humboldt Foundation under Humboldt Professorship, ERC Advanced Grant 339169 "Selfcompletion", by TR 33 "The Dark Universe", and by the DFG cluster of excellence "Origin and Structure of the Universe".

\end{document}